\documentclass[final,3p,times]{elsarticle}
\usepackage{extarrows}
\usepackage{mathrsfs,amsmath,amssymb,amsfonts,latexsym,graphicx,subfigure}

\usepackage{graphics}
\usepackage{graphicx}
\usepackage{epsfig}
\usepackage{psfig}
\usepackage{amssymb}
\usepackage{amsthm}

\journal{arXiv.org}

\newcommand{\beq}{\begin{equation}}
\newcommand{\eeq}{\end{equation}}
\newcommand{\bem}{\begin{displaymath}}
\newcommand{\eem}{\end{displaymath}}
\newcommand{\bey}{\begin{eqnarray}}
\newcommand{\eey}{\end{eqnarray}}

\newcommand{\ud}{\mathrm{d}}
\def\iint{\int\int}

\begin{document}

\begin{frontmatter}

\title{A Symplecticity-preserving Gas-kinetic Scheme for
Hydrodynamic Equations under Gravitational Field}
\author{Jun Luo}
\ead{maluojun@ust.hk}
\author{Kun Xu}
\ead{makxu@ust.hk}
\address{Mathematics Department,  \\
Hong Kong University of Science and Technology  \\
Clear Water Bay, Kowloon, Hong Kong}
\author{Na Liu}
\ead{liuna@lsec.cc.ac.cn}
\address{LSEC, ICMSEC, Academy of Mathematics and Systems Science,\\
Chinese Academy of
Sciences, Beijing 100190, People¡¯s Republic of China}

\begin{abstract}

A well-balanced scheme for a gravitational hydrodynamic system is
defined as a scheme which could precisely preserve a hydrostatic
isothermal solution. In this paper, we will construct a
well-balanced gas-kinetic symplecticity-preserving BGK (SP-BGK)
scheme. In order to develop such a scheme, we model the
gravitational potential as a piecewise step function with a
potential jump at the cell interface. At the same time, the
Liouville's theorem and symplecticity preserving property of a
Hamiltonian flow have been used in the description of  particles
penetration, reflection, and deformation through a potential
barrier. The use of the symplecticity preserving property for a
Hamiltonian flow is crucial in the evaluation of the high-order
moments of a gas distribution function when crossing through a
potential jump. As far as we know, the SP-BGK method is the first
shock capturing Navier-Stokes flow solver with well-balanced
property for a  gravitational hydrodynamic system. A few theorems
will be proved for this scheme, which include the necessity to use
an exact Maxwellian for keeping the hydrostatic state, the total
mass and energy (the sum of kinetic, thermal, and gravitational
ones) conservation, and the well-balanced property to keep a
hydrostatic state during  particle transport and collision
processes. Many numerical examples will be presented to validate the
SP-BGK scheme.

\end{abstract}

\end{frontmatter}

\section{Introduction}\label{sec1}

Generally,
flow equations with source terms can be written as
\begin{equation}\label{eq-source}
U_t+\nabla \cdot F(U)=S,
\end{equation}
where $U$ is the vector of conservative flow variables with corresponding
fluxes $F(U)$ and $S$ is the source term. For a gas flow under an external
time-independent gravitational field, there exists a special solution, i.e.,
the hydrostatic or well-balanced equilibrium solution with a constant
temperature and zero fluid velocity. This solution is an intrinsic solution
due to the balance between the flux gradient and source term, i.e.,
\begin{equation}\label{balance}
\nabla \cdot F(U)=S.
\end{equation}
In order to capture the physical solution for a slowly evolving
gravitational hydrodynamic system, the numerical scheme has to be a well-balanced one
in keeping the hydrostatic solution in the special situation, and has the shock capturing
property in the general case. Theoretically, it seems that to design a well-balanced
shock capturing scheme for the gravitational hydrodynamic system is much more difficult
than that for the shallow water equations.

There have been many attempts to construct
well-balanced gas dynamic codes which  preserve the hydrostatic
solution (\cite{leveque,zingale,botta}).
The schemes in \cite{leveque,zingale,botta} are designed based on the condition
Eq.(\ref{balance}), such as to explicitly  enforce this balance even for the updated non-hydrostatic solution, then use the re-balanced quantities in the evaluation of  fluxes in the next time step. However, for a transient flow, the use of
Eq.(\ref{balance}) directly in the design of the numerical scheme may be problematic, because
in general case Eq.(\ref{balance}) is not satisfied in a physical evolution process,
especially for flow around discontinuities.
So, our aim of this paper is to design a scheme with correct
particle transport and collision across a potential barrier, which will automatically
becomes a well-balanced one when the solution is settling down to the hydrostatic one.
But, the scheme is  still
accurate in capturing any general gas evolution process.

In the past years, a gas-kinetic BGK scheme has been successfully
developed for compressible Euler and Navier-Stokes equations without
gravitational field (\cite{xu1998,xu2001}). The main part of the BGK
scheme is to find a gas distribution function $f$ at a  cell
interface. Physically,  the inclusion of  gravitational effect is
only to change the particle trajectory. Therefore, it should have no
much difficulty for the gas-kinetic scheme to include the
gravitational effect in the modification of the time evolution of a
gas distribution function through the particle acceleration and
deceleration processes. Along this line, the gas kinetic scheme
(GKS) has been extended to a gravitational system \cite{tian}, which
much improved the solution in comparison with operator splitting
method. However, mathematically, the use of a piecewise linear
gravitational potential makes the exact solution complicated and a
simplification of the numerical scheme in \cite{tian} can not keep a
precise well-balanced solution. Therefore, the scheme presented in
\cite{tian} is not a well-balanced one.

In this paper, in order to design a precise well-balanced scheme we
are going to approximate the gravitational potential as a  piecewise
constant function inside each cell with a potential jump at the cell
interface. The detailed particle transport process across a
potential barrier will be followed. In the construction of such a
scheme, the use of the symplecticity property of a Hamiltonian flow
and the Liouville's theorem becomes important in the correct
description of particle penetration, reflection, and deformation
processes across a potential barrier. In a previous paper
\cite{xu-luo}, following the approach of Perthame and Simeoni for
the shallow water equations \cite{perthame}, a well-balanced kinetic
flux vector splitting scheme for gravitational Euler equations has
been developed. However, in the above approach, only a few simple
moments of a gas distribution function are needed, and these simple
moments can be intuitively guessed instead of derived with a solid
physical and mathematical foundation. In order to extend the above
scheme to high-order accuracy and to solve the gravitational NS
equations, a gas-kinetic BGK model with both particle transport and
collision has to be solved. In designing such a scheme, much more
high-order  moments of a gas distribution function have to be
evaluated after the interaction with a potential barrier.
It becomes much harder to construct them intuitively. Furthermore,
to model the particle transport plus collision processes through a
potential barrier is much more challenging than that in the
collision-less case. For example, around a potential jump at a cell
interface,  a multiple equilibrium states have to be constructed on
both sides of a jump. In the construction of such an equilibrium
state for the BGK model, the second law of thermodynamics has to be
satisfied.

The paper is organized as follows. In section \ref{sec:mechanism},
we will present the basic physical principles about the particle
interaction with a potential barrier. The symplectic principle plays
an important role in the design of the well-balanced scheme. Section
\ref{sec:bgk} gives a brief review of the previous BGK scheme
without external forcing field. Section \ref{sec:sp-bgk} presents
particle transport mechanism and the construction of a symplecticity
preserving BGK for the gravitational gas dynamic system. Section
\ref{sec:analysis} is about the theoretical analysis of the schemes,
such as the necessity of using an exact Maxwellian and the
well-balanced property. Section \ref{sec:tests} shows the numerical
tests. The last section is the conclusion.


\section{Particle transport mechanism across a potential barrier}\label{sec:mechanism}

In this paper, the gravitational potential $\phi $ is modeled as a
piecewise constant function. With $\phi_j$ in $jth$-cell and
$\phi_{j+1}$ in $(j+1)th$ cell, there exists a potential jump at
the cell interface, i.e., $\Delta \phi_{j+1/2} = \phi_{j+1} - \phi_j
$. Now what we need to figure out is the effect on  an initial gas
distribution function next to the potential barrier when the
particles move towards the barrier. The associated physical
process could be reflection or penetration of the particles from the
barrier. What we have to evaluate is the relationship between the
moments of the gas distribution functions before and after
interaction with the potential barrier. Since all particles are
located next to the potential jump, the modification of the particle
distribution function happens instantly. Therefore, once a
time-dependent gas distribution function next to the potential
barrier is given, the corresponding distribution after particle
collision with the potential barrier can be evaluated at that
moment. Since the potential jump only affects normal velocity and
its moments, so in this section we only consider distribution
functions with 1-D velocity. The results obtained in this section
will be used in this paper many times on the construction of
symplecticity-preserving scheme.

For an initial gas distribution function $f(u)$ next to a potential
barrier and these particles impacted with the potential jump, the
particle velocity $u$ changes to $u'$, and the distribution
function becomes $\overline{f}(u')$. We are going to use the
following three physical principles to find the relation between
the velocity moments of $\overline{f}(u')$ and $f(u)$.

\noindent {\bf a. Hamiltonian preserving property}: the Hamiltonian
function $H$ of a particle keeps a constant, where
\begin{equation}\label{eq:hamil}
H=\frac{1}{2} u^{2} + \phi(x).
\end{equation}
This is actually the energy conservation for a particle movement
under a conservative potential field. Since we only consider the
interaction of a particle with a potential barrier at an instant
of time,  there are no collisions between particles. Therefore, the
energy conservation for individual particle is precisely conserved, i.e.,
\begin{equation}\label{eq:u-u-rela}
\frac{1}{2} u^{2} + \phi=\frac{1}{2} (u')^{2} + \phi',
\end{equation}
from which the relation between $u$ and $u'$ can be obtained.

\noindent  {\bf b. Liouville's theorem}: the probability density of
a particle in phase space keeps a constant along its movement
trajectory,
\begin{equation}\label{eq:liou}
\overline{f}(u')=f(u).
\end{equation}
In other words, the particle isn't lost or created during its
impact with the potential.

\noindent  {\bf c. The symplecticity preserving property}: for a
Hamiltonian phase flow, we have
\begin{equation}\label{eq:sym}
\iint_{D'}d x'  \ud u' =\iint_Dd x  \ud u,
\end{equation}
where $D'$ and $D$ are the phase volume on the trajectory of the
Hamiltonian phase flow.

During the impact of the particles with the potential barrier,  we
can specially choose $D=(u_1,u_2)\times(u t_1,u t_2)$, then
$D'=(u_1',u_2')\times(u't_1,u' t_2)$ since $D$ and $D'$ are on the
trajectory of the same particle. Therefore,
Eq.(\ref{eq:sym}) goes to
\begin{equation}\label{eq:sym2}
\int_{u_1'}^{u_2'}u'  \ud u' =\int_{u_1}^{u_2}u  \ud u.
\end{equation}
This relationship will be the most important one in the construction
of the moments between between $\overline{f}(u')$ and $f(u)$.
Therefore, the developed scheme in the present paper which uses this
relationship will be called symplecticity-preserving scheme.

With the above three physical principles,  we can derive the
relationship between  the $nth$-order velocity moments of
$\overline{f}(u')$ and that of $f(u)$. From (\ref{eq:liou}) and
(\ref{eq:sym2}), we have
\begin{equation}\label{eq:sym-liou}
\int_{u_1'}^{u_2'}\overline{f}(u')u' \ud u' =\int_{u_1}^{u_2}f(u)u
\ud u
\end{equation}
Moreover, (\ref{eq:hamil}) tells us that $u'$ is a function of $u$,
i.e., $u'=u'(u)$. So, combining with (\ref{eq:sym-liou}), we can get
a general formulation,
\begin{equation}\label{eq:nth}
\textrm{$n$th-order $u$ moment} =\int_{u_1'}^{u_2'}
\overline{f}(u')(u')^{n}\ud u' =\int_{u_1}^{u_2}
f(u)(u'(u))^{n-1}u  \ud u,
\end{equation}
which connects the moments of the distribution functions before and
after impacting with a potential barrier at an instant of time. The
above distribution function can represent the portion of particles
which are reflected or penetrated at the barrier.

\section{A review of gas-kinetic BGK-NS scheme without external forcing field}\label{sec:bgk}

The BGK equation without external forcing field in 2-D is
\begin{equation}
 f_t + \vec{u}\cdot \nabla f= \frac{g-f}{\tau} ,
 \label{eq:bgk}
\end{equation}
where $f$ is the gas distribution function and $g$ is the
equilibrium state approached by $f$, $\nabla f$ is the gradient of
$f$ with respect to $\vec{x}$, $\vec{x}=(x,y)$, and $\vec{u}=(u,v)$
is the particle velocity. The  particle collision time $\tau$ is
related to the viscosity and heat conduction coefficients, i.e.,
$\tau =\mu/p$ where $\mu$ is the dynamic viscosity coefficient and
$p$ is the pressure. The relation between mass $\rho$, momentum $
(\rho U,\,\rho V) $, and energy $\rho E$ densities with the
distribution function $f$ is
\begin{equation}\label{eq:csvt-var}
\begin{pmatrix}{\rho}\\
        {\rho U }\\
        {\rho V}\\
        {\rho E}\\\end{pmatrix} = {\iiint}
    {\psi}
 f \ud u \ud v \ud \xi ,
\end{equation}
where
\begin{displaymath}
\psi =  (\psi_1, \psi_2 , \psi_3, \psi_4)^T = ( 1 , u, v,
{{1\over 2}{( u^2 + v^2 +{ {\xi}}^2) }} )^T ,
\end{displaymath}
$d \xi = d\xi_1 d\xi_2 ...d\xi_K$, and K is the number of degrees of
internal freedom, i.e., $K=(4-2\gamma)/(\gamma-1)$ for 2-D flow.
Since mass, momentum, and energy are conserved during particle
collisions, $f$ and $g$ satisfy the conservation constraint,
\begin{equation}
\iiint (g-f) {\psi}_{\alpha} \ud u \ud v \ud \xi = 0 , \qquad\hbox{} \qquad {\alpha
= 1,2,3,4}\hbox{} \qquad\hbox{}
 \label{eq:constraint}
\end{equation}
at any point in space and time.
The integral solution of (\ref{eq:bgk}) is
\begin{equation}\label{eq:dis}
f(\vec{x},t,\vec{u},\xi) = {1\over \tau} \int_0^t g (\vec{x'},t',\vec{u},\xi)
e^{-(t-t')/{\tau}} dt'
   +e^{-t/{\tau}} f_0(\vec{x}-\vec{u} t,\vec{u},\xi) ,
\end{equation}
where $\vec{x'}=\vec{x}-\vec{u}( t-t')$ is the particle trajectory.
The solution $f$ in (\ref{eq:dis}) solely depends on the modeling of
$f_0$ and $g$.

For a finite volume scheme, we need to evaluate the fluxes across a
 cell interface in order to update the cell averaged conservative flow variables. In
the BGK scheme, the fluxes are defined by
\begin{equation}\label{eq:flux}
\begin{pmatrix}{F_{\rho}}\\
        {F_{\rho U}}\\
        {F_{\rho V}}\\
        {F_{\rho E}}\\\end{pmatrix} = {\iiint}
    u{\psi}
 f \ud u \ud v \ud \xi ,
\end{equation}
which depends on the gas distribution function $f$ in
Eq.(\ref{eq:dis}) at the cell interface. Let's consider the
construction of the distribution function at the cell interface
$\vec{x}_{j+1/2}=(x_{j+1/2},y_i)$, where $\vec{x}_{j+1/2}$ is the
location of the cell interface center in the physical domain.
Locally, around this cell interface, with the assumption of the
x-direction as the normal direction and y-direction as the
tangential direction, based on the BGK model a solution in this
local coordinate can be obtained.

\begin{figure}
\begin{center}
\includegraphics[scale=0.3,bb = 0 420 600 720, clip=true]{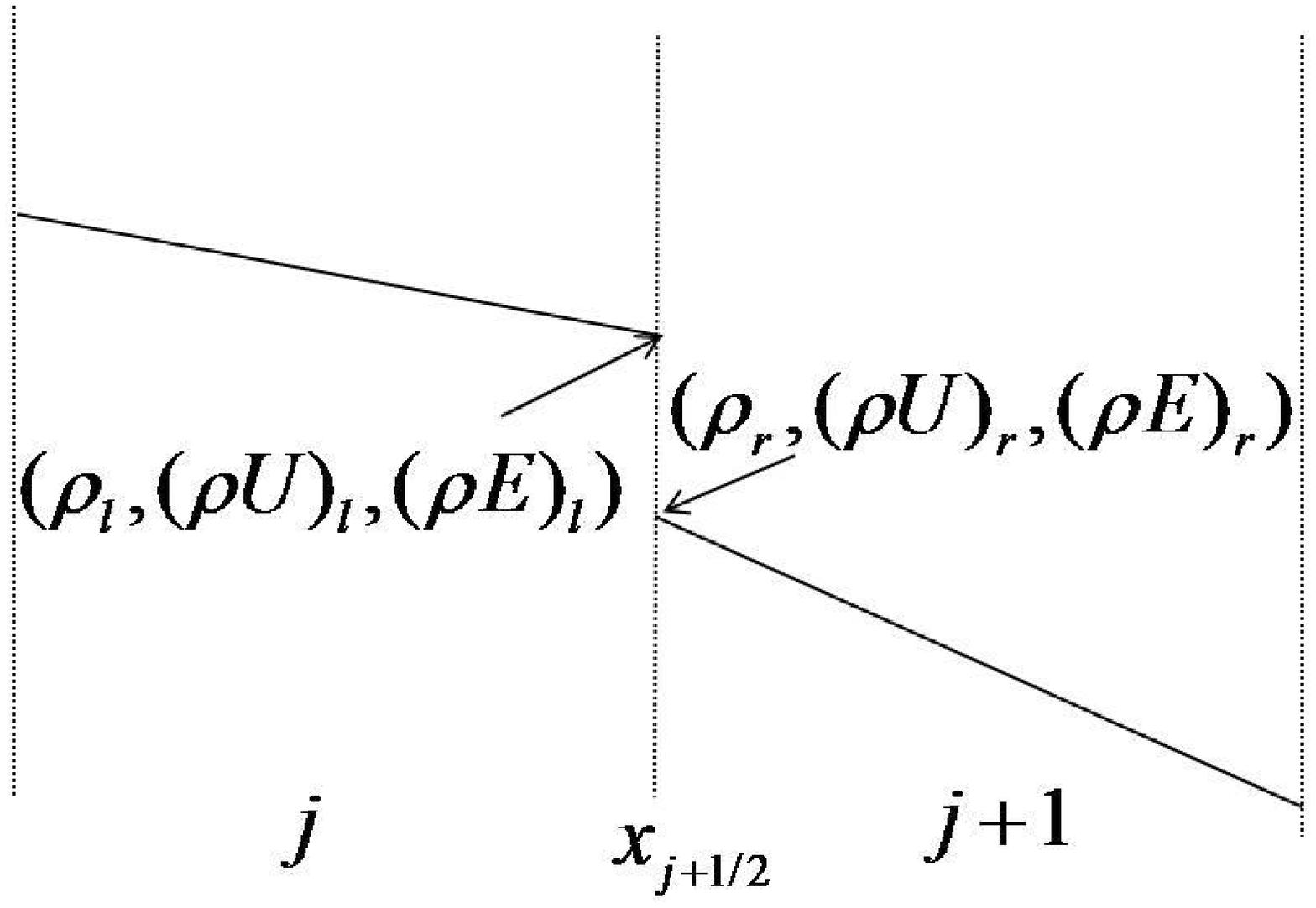}
\caption{\small Reconstruction of the conservative variables at the cell interface.}\label{fig:recon}
\end{center}
\end{figure}

\begin{figure}
\begin{center}
\includegraphics[scale=0.25,bb = 40 360 700 790, clip=true]{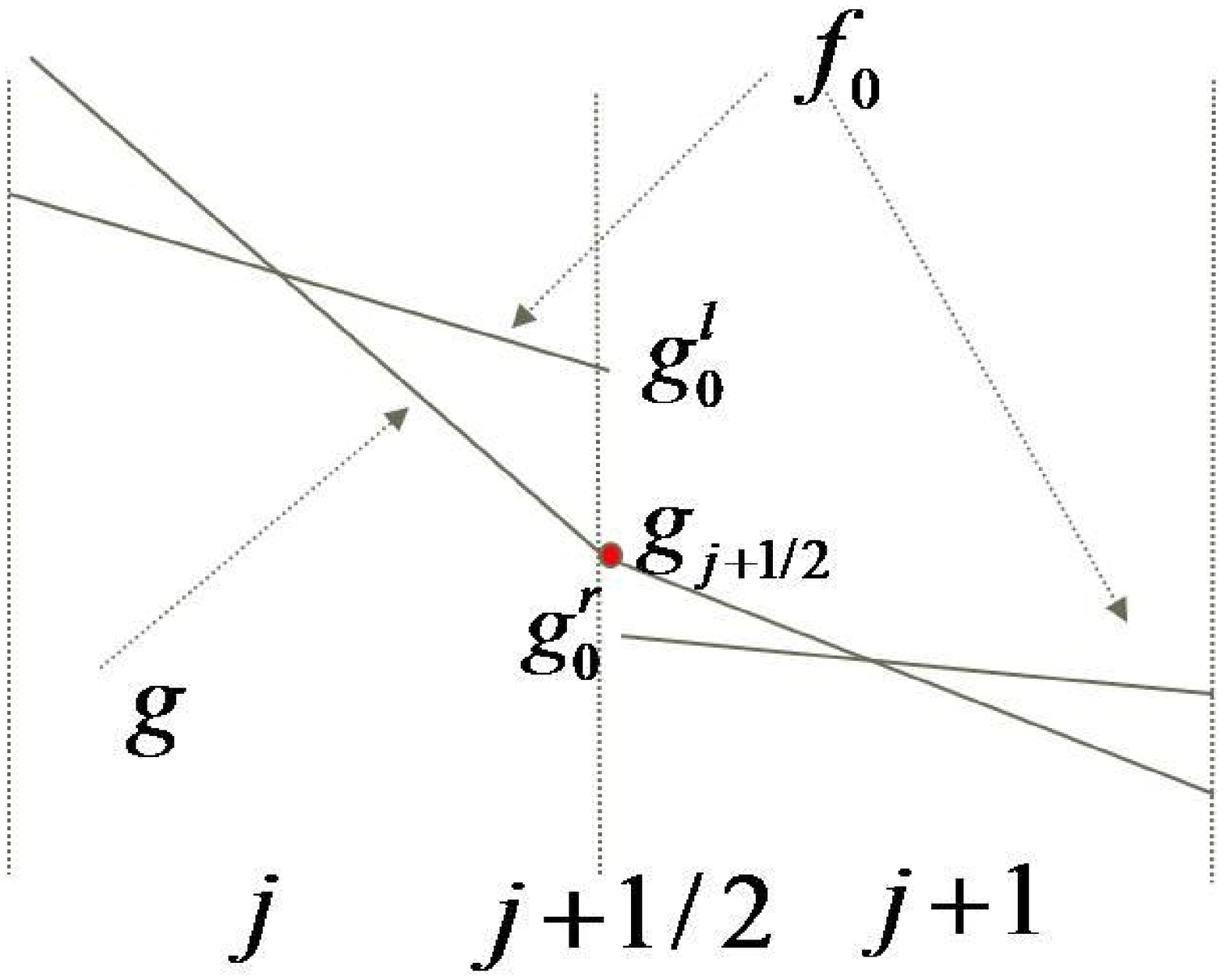}
\caption{The modeling of the initial and equilibrium distribution functions at the cell interface for the BGK scheme without external forcing field.} \label{fig:bgk}
\end{center}
\end{figure}

By using the MUSCL-type limiter, a discontinuous reconstruction of
 the macroscopic flow variables can be obtained around the cell
interface (see fig.\ref{fig:recon}). The initial gas distribution
function $f_0$ in (\ref{eq:dis}) on both sides of a cell interface
can be constructed as
\begin{equation}\label{eq:f0}
\begin{array}{ll}
f_0^l(\vec{x},\vec{u},\xi)=g_0^l(1+a^l (x-x_{j+1/2})+b^l(y-y_i)-\tau(a^l u+b^l v+A^l)),& x \le x_{j+1/2},\\ \\
f_0^r(\vec{x},\vec{u},\xi)=g_0^r(1+a^r (x-x_{j+1/2})+b^r(y-y_i)-\tau(a^r u+b^r v+A^r)),& x>x_{j+1/2},
\end{array}
\end{equation}
where the Chapman-Enskog expansion up to the Navier-Stokes order has
been used in the above initial reconstruction. Here $g_0^l$ and
$g_0^r$ are the corresponding Maxwellians to $W^l=(\rho_l , (\rho
U)_l, (\rho V)_l, (\rho E)_l)$ and $W^r=(\rho_r , (\rho U)_r, (\rho
V)_r, (\rho E)_r)$ at both sides of the interface. The  Maxwellian
distribution function corresponding to $W=(\rho, (\rho U), (\rho V),
(\rho E))$ has the form
\begin{equation}\label{eq:max}
g=\rho \left(\frac{\lambda}{\pi}\right)^{\frac{K+2}{2}}
e^{\lambda ((u-U)^2+(v-V)^2+\xi^2)},
\end{equation}
where $\lambda$ is equal to $ m/ 2 kT$, $m$ is the
molecular mass, $k$ is the Boltzmann constant, and $T$ is the
temperature.
The equilibrium distribution
functions around the cell interface can be modeled as
\begin{equation}\label{eq:g}
\begin{array}{ll}
g^l(\vec{x},t,\vec{u},\xi)=g_{j+1/2}^l(1+\overline{a}^l (x-x_{j+1/2})+\overline{b}^l (y-y_i)+\overline{A}^l t),& x \le x_{j+1/2},\\ \\
g^r(\vec{x},t,\vec{u},\xi)=g_{j+1/2}^r(1+\overline{a}^r (x-x_{j+1/2})+\overline{b}^r (y-y_i)+\overline{A}^r t),& x>x_{j+1/2}.
\end{array}
\end{equation}
In the case without external forcing term, $g_{j+1/2}^l$ and
$g_{j+1/2}^r$ in the above equation are the same distribution
functions, i.e., $g_{j+1/2}^l=g_{j+1/2}^r$ (see fig.\ref{fig:bgk}),
which can be obtained using the conservation constraint (\ref{eq:constraint})
at $\vec{x}=\vec{x}_{j+1/2}$ and $t\rightarrow 0$,
\begin{equation}\label{eq:g-inte1}
\begin{array}{l}
\quad\iiint g_{j+1/2}^l \psi \ud u \ud v \ud \xi=
\iiint g_{j+1/2}^r \psi \ud u \ud v \ud \xi=
W_{j+1/2}\\ \\=
\iiint_{u>0} f_0^l(\vec{x}_{j+1/2},\vec{u},\xi) \psi \ud u \ud v \ud \xi+\iiint_{u<0} f_0^r(\vec{x}_{j+1/2},\vec{u},\xi) \psi \ud u \ud v \ud \xi.
\end{array}
\end{equation}
Therefore, at the cell interface the final distribution function can
be fully determined using the integral solution (\ref{eq:dis}). The
final distribution function can be written as
\begin{equation}\label{eq:dis-bgk}
\begin{array}{l}
\quad f(\vec{x}_{j+1/2},t,\vec{u},\xi)\\ \\=\left\{
\begin{array}{l}
f^l(\vec{x}_{j+1/2},t,\vec{u},\xi)\quad\quad u \ge 0, \\ \\
f^r(\vec{x}_{j+1/2},t,\vec{u},\xi)\quad\quad u< 0,
\end{array}\right.\\ \\
=\left\{
\begin{array}{l}
\frac{1} {\tau} \int_0^t g^l (\vec{x}_{j+1/2}-\vec{u}(t-t'),t',\vec{u},\xi)
e^{-(t-t')/{\tau}} dt'+ e^{-t/{\tau}} {f_{0}^l} (\vec{x}_{j+1/2}-\vec{u} t) , \quad\quad
u \ge 0,\\ \\
\frac{1} {\tau} \int_0^t g^r
(\vec{x}_{j+1/2}-\vec{u}(t-t'),t',\vec{u},\xi) e^{-(t-t')/{\tau}}
dt'+ e^{-t/{\tau}} {f_{0}^r} (\vec{x}_{j+1/2}-\vec{u} t) ,
\quad\quad u< 0,
\end{array}\right.
\end{array}
\end{equation}
which can be used to evaluate the fluxes
\begin{equation}\label{eq:flux-withoutG}
\begin{array}{l}
\quad F_{j+1/2}^l(t)=F_{j+1/2}^r(t)\\ \\=\iiint_{u>0}
uf^l(\vec{x}_{j+1/2},t,\vec{u},\xi) \psi \ud u \ud v \ud
\xi+\iiint_{u<0} uf^r(\vec{x}_{j+1/2},t,\vec{u},\xi) \psi \ud u \ud
v \ud \xi.
\end{array}
\end{equation}
The update of the cell averaged conservative variables becomes
\begin{equation}\label{fv}
W^{n+1}_j=W^n_j+\frac{1}{\Delta x} \int^{t_{n+1}}_{t_n}
\left[F^r_{j-1/2}(t)-F^l_{j+1/2}(t)\right]  dt +\frac{1}{\Delta y}
\int^{t_{n+1}}_{t_n} \left[F^r_{i-1/2}(t)-F^l_{i+1/2}(t)\right]  dt,
\end{equation}
where $F^l_{j-1/2}(t)$ ... $F^r_{i+1/2}(t)$ are the fluxes at the
center of the cell interfaces.

The definitions and constructions of all parameters related to the
spatial and temporal slopes, such as $a$, $b$ and $A$, can be found
in \cite{xu1998} and \cite{xu2001}.

In summary, at the cell interface $\vec{x}_{j+1/2}$ we can construct
the equilibrium distribution functions $g_{j+1/2}^l$ and
$g_{j+1/2}^r$ from initial distribution $f_0^l$ and $f_0^r$. Also,
we can find fluxes $F_{j+1/2}^l(t)$ and $F_{j+1/2}^r(t)$ from the
integral solution $f^l$ and $f^r$. Without external forcing field, all
the particles running into the cell interface can freely cross it.
Therefore, the equilibrium states and fluxes at the interface have
unique values, i.e., $g_{j+1/2}^l=g_{j+1/2}^r$ and
$F_{j+1/2}^l(t)=F_{j+1/2}^r(t)$. However, with the approximation of
constant potential inside each cell and a potential jump at the cell
interface, the modeling of equilibrium state $g$ around a cell
interface has to be considered separately on different sides of the
cell interface, where $g_{j+1/2}^l \neq g_{j+1/2}^r$ in general
case. But, the mathematical formulae described in (\ref{eq:g}) and
the integral solution in Eq.(\ref{eq:dis-bgk}) can be still used.
One of the main reason for the validity of the integral solution is
that there is no gravitational force inside each cell. However, the
construction of the equilibrium states and the calculation of fluxes
will not be as simple as that in (\ref{eq:g-inte1}) and
(\ref{eq:flux-withoutG}). In the evaluation of the equilibrium
states and the fluxes, the physical principles for the particle
transport discussed in the last section have to be used. In the next
section, the determination of $g$ and fluxes will be described.


\section{The symplecticity preserving BGK(SP-BGK) scheme}\label{sec:sp-bgk}

In this section, we will construct a well-balanced gas-kinetic
scheme for hydrodynamic equations under gravitational field. In
order to clarify the concepts, we are going to use a similar
procedure as that of the construction of the BGK-NS scheme without
external forcing field.

\subsection{The initial data reconstruction}

For a hydrostatic solution, the flow variables satisfy the
conditions,
\begin{equation}\label{eq:recon}
U=0,\,V=0,\,\lambda=constant,\,Ba=constant,
\end{equation}
where $Ba=\rho e^{2\lambda \phi}$. In order to avoid introducing
errors in the initial reconstruction for the hydrostatic case, it is
reasonable to use the variables $(U,\,V,\,\lambda,\,Ba)$ in the
reconstruction. More specifically,  we firstly apply a MUSCL-type
limiter to reconstruct the slopes of $(U,\,V,\,\lambda,\,Ba)$, i.e.,
$(S_U,\,S_V,\,S_{\lambda},\,S_{Ba})$ inside each cell. Since
\[
\rho=\frac{Ba}{e^{2\lambda \phi}},\,\rho E=\frac{1}{2} \rho (U^2+V^2)+\frac{K+2}{4\lambda}\rho,
\]
we can get the corresponding slopes for other flow variables, \\
\[S_{\rho}=\frac{1}{e^{2\lambda \phi}}S_{Ba}-2\rho \phi S_{\lambda},
S_{\rho U}=S_{\rho}U +\rho S_U,
S_{\rho V}=S_{\rho}V +\rho S_V,\]\\
\[S_{\rho E}=\left[\frac{1}{2} (U^2+V^2)+\frac{K+2}{4\lambda}\right]S_{\rho}
+\rho \left[U S_U+ V S_V-\frac{K+2}{4\lambda^2}S_{\lambda}\right],\]
where $(S_{\rho},\,S_{\rho U},\,S_{\rho V},\,S_{\rho E})$ are the
slopes of $(\rho,\,\rho U,\,\rho V,\,\rho E)$ inside that cell.
Therefore, we can reconstruct $(\rho,\,\rho U,\,\rho V,\,\rho E)$ in
each cell using their cell averaged quantities and the above slopes.
Here, all slopes become zeros when the initial flow is in a
hydrostatic state, and the reconstruction will not introduce
numerical errors.  In the general case, the above reconstruction
works as well.

\subsection{The gas-kinetic SP-BGK scheme}

\begin{figure}[h]
\centering
\includegraphics[angle=90,scale=0.17,bb = 40 60 480 800, clip=true]{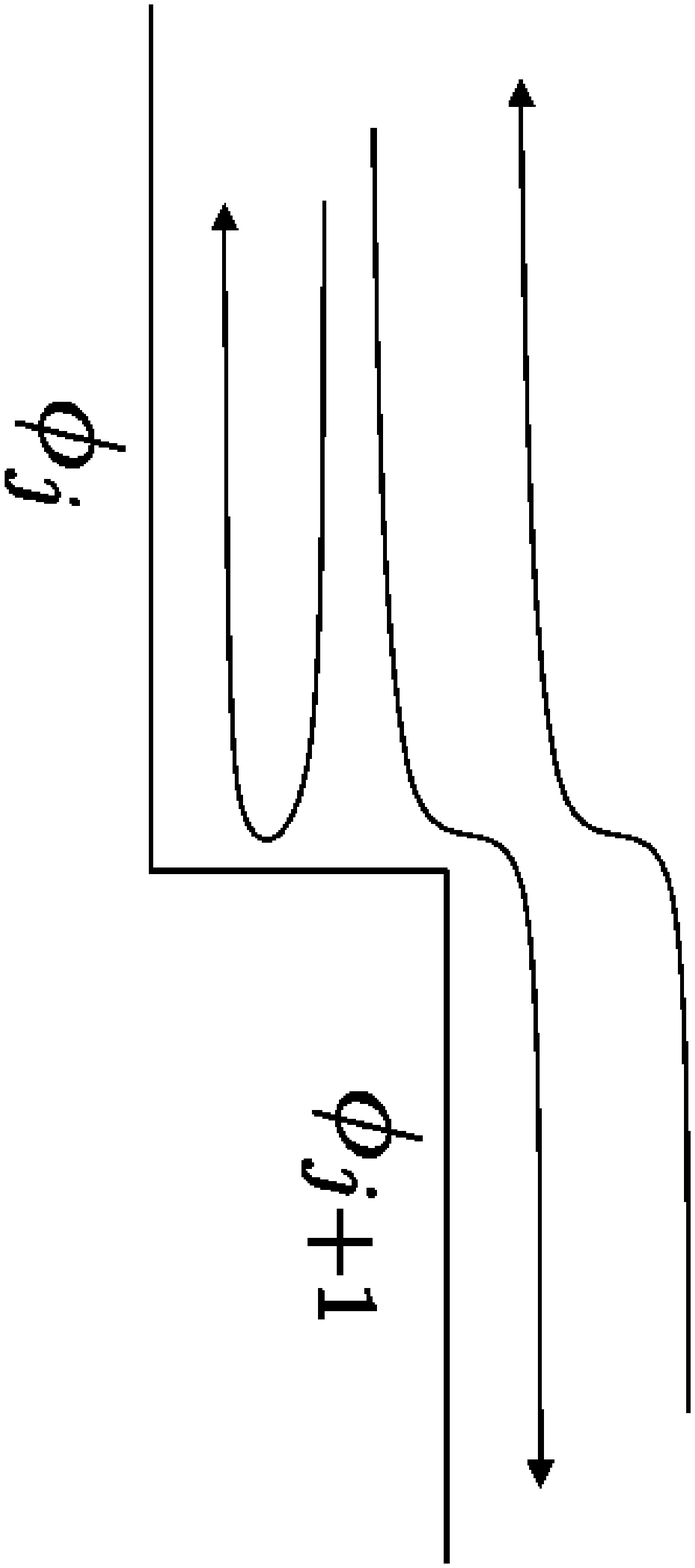}
\caption{\small The particles' movement at the interface with
potential jump $\phi_j<\phi_{j+1}$. } \label{crossing}
\end{figure}

\begin{figure}
\begin{center}
\includegraphics[scale=0.2,bb = 2 -30 800 460, clip=true]{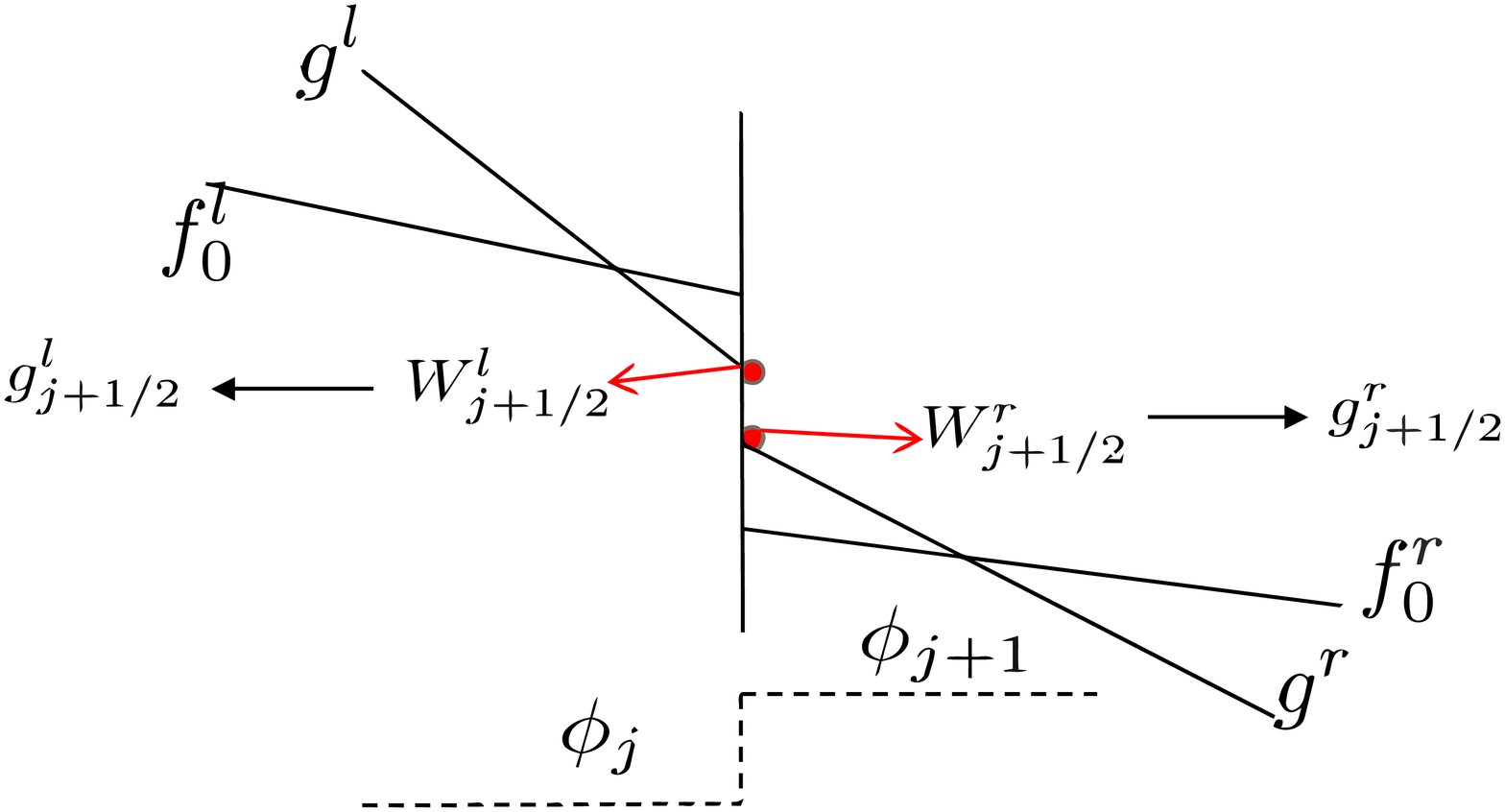}
\caption{The modeling of the initial and equilibrium distribution
functions at the cell interface for the SP-BGK scheme with a
potential jump at the cell interface.} \label{fig:sp-bgk}
\end{center}
\end{figure}

With the modeling of piecewise constant gravitational potential
inside each cell, i.e., $\phi_j$ inside the $jth$ cell, there is a
potential jump at the cell interface $\vec{x}_{j+1/2}$. It is
obvious that the distribution function $f$ also satisfies the
equation (\ref{eq:bgk}) inside each cell since there is no external
forcing term inside each cell. Therefore, the similar framework used
in the constructing BGK-NS scheme can be extended here to design the
SP-BGK scheme with gravitational field. For example, with the
initial reconstruction, the non-equilibrium states around each cell
interface can be obtained. Also, due to the potential jump, the
equilibrium states are different in the left and right hand sides of
the interface, but the integral solution of the BGK model can be
still used in the construction of the local solution separately
around the cell interface. However, at the cell interface, we have
to consider the effect of the potential jump on the particle
movement. Since the equilibrium states, $g_{j+1/2}^l$ and
$g_{j+1/2}^r$, and the fluxes, $F_{j+1/2}^l(t)$ and
$F_{j+1/2}^r(t)$, involve the particle interaction with the
potential jump, we will show that $g_{j+1/2}^l \neq g_{j+1/2}^r$ in
Eq.(\ref{eq:g})(see fig.\ref{fig:sp-bgk}), and $F_{j+1/2}^l(t)\neq
F_{j+1/2}^r(t)$ in the general case. Their determination depends on
the particle transport modeling. The potential jump gives a critical
speed $U_c=\sqrt{2|\phi_j-\phi_{j+1}|}$, which provides a threshold
for the particle movement. Because of the potential jump, not all
particles running into the cell interface could go through freely.
Some may be reflected due to less kinetic energy to overcome the
potential barrier (see fig.\ref{crossing}). For these particles
passing  through the cell interface, their momentum and energy need
to be modified due to particle acceleration during the transport
process.

Without losing generality, we only discuss the case of
$\phi_j<\phi_{j+1}$ in this subsection. Using similar methods and
ideas, all the formulae for the case $\phi_j>\phi_{j+1}$ can  be
easily obtained. Let's assume the initial reconstructed gas
distribution at a cell interface before the interaction with the
potential jump is
\begin{equation}
f(\vec{x}_{j+1/2},t,\vec{u},\xi)=\left\{\begin{array}{ll}f_j(\vec{x}_{j+1/2},t,\vec{u},\xi), & u \ge 0,\\ \\
f_{j+1}(\vec{x}_{j+1/2},t,\vec{u},\xi), & u<0 . \end{array}\right.
\end{equation}

Starting from the above distribution function, the particle
collision with the potential jump changes distribution functions
to $f_{j+1/2}^l(t,\vec{u},\xi)$ and $f_{j+1/2}^r(t,\vec{u},\xi)$ at
the left and right hand sides of the cell interface respectively,
which can be represented as
\begin{equation}\label{dis-left}
f_{j+1/2}^l(t,\vec{u},\xi)=\left\{
\begin{array}{ll}
f_j(\vec{x}_{j+1/2},t,\vec{u},\xi) , & u>0,\\ \\
\tilde{f}_j(\vec{x}_{j+1/2},t,\vec{u},\xi) , & 0 \ge u>-U_c,\\ \\
\overline{f}_{j+1}(\vec{x}_{j+1/2},t,\vec{u},\xi) , & u \le -U_c,
\end{array}
\right.
\end{equation}
and
\begin{equation}\label{dis-right}
f_{j+1/2}^r(t,\vec{u},\xi)=\left\{\begin{array}{ll}
\overline{f}_j(\vec{x}_{j+1/2},t,\vec{u},\xi), & u \ge 0,\\ \\
f_{j+1}(\vec{x}_{j+1/2},t,\vec{u},\xi),  & u<0.
\end{array}
\right.
\end{equation}
The definition of the above distribution functions is from the
following physical consideration (see fig.\ref{crossing}). Because
the potential jump is only at the normal direction of the cell
interface,  it only affects the normal particle velocity, $u$. In
(\ref{dis-left}), $\tilde{f}_j$ is the distribution function of the
reflected particle in the $jth$ cell with the original distribution
function $f_j$ which has a positive particle velocity less than
$U_c$. Here $\overline{f}_{j+1}$ is the distribution function of the
particle in the $jth$ cell coming from the $(j+1)th$ cell with the
original distribution function $f_{j+1}$ with negative particle
velocity. This particle has been accelerated in the negative normal
direction after passing through the cell interface. Also,
$\overline{f}_j$ is the distribution function of the particle in the
$(j+1)th$ cell coming from the $jth$ cell with the original
distribution function $f_j$ and positive velocity higher than $U_c$.
This particle has been be decelerated in the positive normal
direction after passing through the cell interface. Therefore, the
effect of the potential jump modifies the distribution function, but
the particle velocity moments of the modified distribution function
and the original ones
 are related through the physical
principles which have been introduced in section
\ref{sec:mechanism}.

Here, we will show the procedure of the SP-BGK scheme first, then
clarify the detailed derivation of the formulae for equilibrium
states and fluxes.

Using particle free transport mechanism in Eq.(\ref{eq:dis}) for the
initial gas distribution function $f_0$, i.e.,
$f_j(\vec{x}_{j+1/2},t,\vec{u},\xi)=f_0^l(\vec{x}_{j+1/2}-\vec{u}t,\vec{u},\xi)$
and
$f_{j+1}(\vec{x}_{j+1/2},t,\vec{u},\xi)=f_0^r(\vec{x}_{j+1/2}-\vec{u}t,\vec{u},\xi)$,
and due to their interaction with the potential jump, the initial
condition will be changed according to Eq.(\ref{dis-left}) and
(\ref{dis-right}), from which two sets of conservative variables at
different sides of the cell interface can be obtained,
\begin{equation}\label{eq:equi-l}
\begin{array}{l}
W_{j+1/2}^l=\iiint_{-\infty}^{\infty}f_{j+1/2}^l(t=0,\vec{u},\xi) \psi \ud u \ud v \xi\\ \\
\quad \quad \quad=\iiint_{0}^{+\infty} f_j(\vec{x}_{j+1/2},t=0,\vec{u},\xi) \psi \ud u
 +\iiint_{-U_c}^0  \tilde{f}_j(\vec{x}_{j+1/2},t=0,\vec{u},\xi)\psi \ud u \ud v \ud \xi\\ \\
\qquad \quad \quad+\iiint_{-\infty}^{-U_c} \overline{f}_{j+1}(\vec{x}_{j+1/2},t=0,\vec{u},\xi)
 \psi
\ud u \ud v \ud \xi,
\end{array}
\end{equation}\\
and\\
\begin{equation}\label{eq:equi-r}
\begin{array}{l}
W_{j+1/2}^r=\iiint_{-\infty}^{\infty}f_{j+1/2}^r(t=0,\vec{u},\xi) \psi \ud u \ud v \xi\\ \\
\quad \quad
\quad=\iiint_0^{+\infty}\overline{f}_j(\vec{x}_{j+1/2},t=0,\vec{u},\xi)
\psi \ud u \ud v \ud \xi +\iiint_{-\infty}^0
f_{j+1}(\vec{x}_{j+1/2},t=0,\vec{u},\xi) \psi \ud u \ud v \ud \xi,
\end{array}
\end{equation}
from which, two Maxwellians $g_{j+1/2}^l$ and $g_{j+1/2}^r$ in the
equilibrium states (\ref{eq:g}) can be fully determined. Then,
following the method used in the development of BGK-NS scheme
\cite{xu2001}, the final gas distribution at the left and right hand
sides of a cell interface, i.e., $f^l$ and $f^r$ in
(\ref{eq:dis-bgk}), can be obtained. When choosing the integral
solutions as the original distribution functions, i.e.,
$f_j(\vec{x}_{j+1/2},t,\vec{u},\xi)=f^l(\vec{x}_{j+1/2},t,\vec{u},\xi)$
and
$f_{j+1}(\vec{x}_{j+1/2},t,\vec{u},\xi)=f^r(\vec{x}_{j+1/2},t,\vec{u},\xi)$,
and considering their interactions with the potential jump, these
distribution functions will be modified as Eq.(\ref{dis-left}) and
(\ref{dis-right}), from which the corresponding fluxes at different
sides of the cell interface can be determined,
\begin{equation}\label{flux-l}
\begin{array}{l}
F_{j+1/2}^l(t)=\iiint_{-\infty}^{+\infty}u f_{j+1/2}^l(t,\vec{u},\xi) \psi \ud u \ud v \ud \xi\\ \\
\qquad \qquad=\iiint_{0}^{+\infty}u f_j(\vec{x}_{j+1/2},t,\vec{u},\xi) \psi \ud u+\iint_{-U_c}^0 u \tilde{f}_j(\vec{x}_{j+1/2},t,\vec{u},\xi)
\psi \ud u \ud v \ud \xi\\ \\
\qquad \qquad \quad+\iiint_{-\infty}^{-U_c}u
\overline{f}_{j+1}(\vec{x}_{j+1/2},t,\vec{u},\xi) \psi \ud u \ud v
\ud \xi,
\end{array}
\end{equation}
and
\begin{equation}\label{flux-r}
\begin{array}{l}
F_{j+1/2}^r(t)=\iiint_{-\infty}^{+\infty}u f_{j+1/2}^r(t,\vec{u},\xi) \psi \ud u \ud v \ud \xi\\ \\
\qquad \qquad=\iiint_0^{+\infty}u \overline{f}_j(\vec{x}_{j+1/2},t,\vec{u},\xi) \psi \ud u \ud v \ud \xi
+\iiint_{-\infty}^0 u f_{j+1}(\vec{x}_{j+1/2},t,\vec{u},\xi) \psi \ud u \ud v \ud \xi.
\end{array}
\end{equation}
Note that due to the potential jump, in general we have
$g_{j+1/2}^l\ne g_{j+1/2}^r$ and $F_{j+1/2}^l\ne F_{j+1/2}^r$.
Finally, we can use (\ref{fv}) to update the cell averaged
conservative variables.

In the above formulae (\ref{eq:equi-l}), (\ref{eq:equi-r}),
(\ref{flux-l}) and (\ref{flux-r}), we need to find the $nth$ order
velocity moments of the modified distribution functions,
$\tilde{f}_j$, $\overline{f}_{j+1}$ and $\overline{f}_j$, which can
be evaluated from the moments of the original distribution funcions
$f_j$, $f_{j+1}$ and $f_j$ respectively by (\ref{eq:nth}). Let's
figure out how to evaluate the $nth$ order normal velocity moments
of $\tilde{f}_j(u)$, $\overline{f}_{j+1}(u)$ and
$\overline{f}_j(u)$.

\noindent a. {\bf The $nth$-order normal velocity moments of
$\tilde{f}_j$}

Recall that $\tilde{f}_j$ is the distribution function of the
reflected particle in the $jth$ cell. Assume that the normal
particle velocity is $u$ before the reflection, and the distribution
of the particle before reflection is $f_j(u)$ with $0<u<U_c$. After
the reflection, its velocity becomes $u'$ and $u'=-u$, for these
particles, (\ref{eq:nth}) gives
\begin{equation}\label{eq:reflection}
\int_{-U_c}^0 \tilde{f}_j(u') (u')^n\ud u'
=\int_{U_c}^0 f_j(u) u(-u)^{n-1}\ud u
=\int_0^{U_c} f_j(u)(-1)^n u^n  \ud u.
\end{equation}

\noindent b. {\bf The $nth$-order normal velocity moments of
$\overline{f}_{j+1}$}

$\overline{f}_{j+1}$ is the distribution function of the particle in
the $jth$ cell  coming from the $(j+1)th$ cell. Its distribution
function before crossing the potential jump is $f_{j+1}$ with
normal velocity $u<0$. After passing through the interface, the
normal velocity changes from $u$ to $u'$, where $u$ and $u'$ are
related by the Hamiltonian preserving property, i.e.,
\[
\frac{1}{2}u^2+\phi_{j+1}=\frac{1}{2}(u')^2+\phi_{j}.
\]
So, $u'=-\sqrt{u^2+U_c^2}$, Eq.(\ref{eq:nth}) gives
\begin{equation}\label{eq:cross1}
\int_{-\infty}^{-U_c} \overline{f}_{j+1}(u')(u')^n \ud u'
=\int_{-\infty}^0 f_{j+1}(u)(-1)^{n-1}u (u^2+U_c^2)^{(n-1)/2} \ud u.
\end{equation}

\noindent c. {\bf The $nth$-order normal velocity moments of
$\overline{f}_{j}$}

$\overline{f}_{j}$ is the distribution function of the particle in
the $(j+1)th$ cell coming from the $jth$ cell. Its distribution
function before passing through the potential jump is $f_j$ with
normal velocity $u>U_c$. After passing through the cell interface,
the normal velocity changes to $u'$. The relation between $u$ and
$u'$ becomes
\[
\frac{1}{2}u^2+\phi_{j}=\frac{1}{2}(u')^2+\phi_{j+1}.
\]
So, $u'=\sqrt{(u)^2-U_c^2}$, Eq.(\ref{eq:nth}) deduces
\begin{equation}\label{eq:cross2}
\int_0^{+\infty}\overline{f}_j(u') (u')^n\ud u
=\int_{U_c}^{+\infty} f_j(u)u (u^2-U_c^2)^{(n-1)/2}\ud u.
\end{equation}

Based on the above moment evaluations, we can get the formulae for
$W_{j+1/2}^l$, $W_{j+1/2}^r$, $F_{j+1/2}^l(t)$ and $F_{j+1/2}^r(t)$
by (\ref{eq:equi-l})- (\ref{eq:cross2}) for the case
$\phi_j<\phi_{j+1}$. The formulae for the case $\phi_j>\phi_{j+1}$
can be found similarly. All the formulae are given in the appendix
for both 1-D and 2-D cases. Therefore, the SP-BGK scheme is
presented.


\subsection{Limiting Cases}

\noindent a. {\bf The 1st order SP-BGK scheme}

When all the slopes in the reconstruction are zeros, and all slopes
 $a$, $b$ and $A$ of the distribution function in (\ref{eq:f0}) and (\ref{eq:g}) become zeros,
the SP-BGK scheme becomes a 1st order scheme. Now, the distribution
function in (\ref{eq:dis}) becomes
\[
f(\vec{x}_{j+1/2},t,\vec{u},\xi)=\left\{ \begin{array}{ll}
(1-e^{-t/\tau}) g_{j+1/2}^l+e^{-t/\tau} g_0^l, & u \ge 0,\\ \\
(1-e^{-t/\tau}) g_{j+1/2}^r+e^{-t/\tau} g_0^r, &
u<0.
\end{array} \right.
\]
Or, with the definition of a small parameter $\varepsilon$, i.e.,
$0<\varepsilon<1$, the distribution function becomes
\begin{equation}
f(\vec{x}_{j+1/2},t,\vec{u},\xi)=\left\{ \begin{array}{ll}
(1-\varepsilon) g_{j+1/2}^l + \varepsilon g_0^l, & u \ge 0,\\ \\
(1-\varepsilon) g_{j+1/2}^r + \varepsilon g_0^r, &
u<0,
\end{array} \right.
\end{equation}
which is called the 1st-order SP-BGK scheme.\\ \\

\noindent b. {\bf The SP-KFVS scheme}

When the collision time $\tau$ goes  to $+\infty$, the distribution
function in (\ref{eq:dis-bgk}) becomes
\begin{equation}\label{eq:dis-kfvs}
\begin{array}{l}
\quad f(\vec{x}_{j+1/2},t,\vec{u},\xi)=\left\{
\begin{array}{l}
f^l(\vec{x}_{j+1/2},t,\vec{u},\xi)\quad\quad u \ge 0, \\ \\
f^r(\vec{x}_{j+1/2},t,\vec{u},\xi)\quad\quad u< 0,
\end{array}\right. \\ \\
\qquad\qquad\qquad\qquad=\left\{
\begin{array}{l}
{f_{0}^l} (\vec{x}_{j+1/2}-\vec{u} t) , \quad\quad
u \ge 0,\\ \\
{f_{0}^r} (\vec{x}_{j+1/2}-\vec{u} t) , \quad\quad
u< 0.
\end{array}\right.
\end{array}
\end{equation}
The above solution solely comes from free transport and there is no
contribution of the equilibrium states $g$ in the integral solution
$f$. It equals to solve
\[
f_t + \vec{u}\cdot \nabla f=0
\]
directly when the initial distribution function is modeled as
(\ref{eq:f0}). In other words, we don't consider particle collision
here, and  needn't to model the equilibrium distribution function
$g$ in (\ref{eq:g}). This is exactly the same scheme introduced in
\cite{xu-luo}, which is called SP-KFVS scheme.
It is actually a limiting case of the SP-BGK scheme.\\

In this section, with the assumption of piecewise constant
gravitational potential, a SP-BGK scheme is presented. As will be
presented in the next section, the SP-BGK scheme is a well-balanced
scheme for the gravitational hydrodynamic system. This is the first
well-balanced scheme, which has the shock capturing property as well
in the general case.


\section{Theoretical analysis}\label{sec:analysis}
For simplicity, we are going to prove all the theorems in the 1-D
case. But all the conclusions still  hold for higher dimensions as
well, because there is no dynamic difference in higher dimensions
when the potential jump is modeled as a piecewise constant function. \\

In the current scheme, the updated flow variables inside each cell
are the mass, momentum, and energy densities (kinetic + thermal
ones). The gravitational energy is not explicitly included. However,
for an isolated gravitational system, the total energy (kinetic +
thermal + gravitational ones) conservation is a necessary condition
in order to get a correct physical solution. In the following
theorem, we are going first to prove that the conservation of total
energy in the current kinetic scheme is satisfied.

\noindent{\bf Theorem 3.1:} The SP-KFVS and SP-BGK schemes are mass
and total energy conservative schemes.

\noindent {\bf Proof} \quad The only difference between the SP-KFVS
and SP-BGK schemes is that they have different original distribution
functions $f_j(u)$ and $f_{j+1}(u)$. However, whatever $f_j(u)$ and
$f_{j+1}(u)$ are, the mass and total energy are conserved when the
fluxes are calculated by (\ref{flux1-l}) and (\ref{flux1-r}) or
(\ref{flux2-l}) and (\ref{flux2-r}) in the appendix. The concept of
conservation of a variable means that the change of that variable in
any fixed domain depends only on the fluxes across the interfaces of
that control volume. In the following proof, we assume the control
volume consists of many cells between the cell index $K_1$ and
$K_2$, where $K_1 <  K_2$. Then, we need to prove that the change of
the mass and total energy in the control volume depends only on the
fluxes at the interfaces $x_{K_1-1/2}$ and $x_{K_2+1/2}$. Without
losing generality, we assume $\phi_j<\phi_{j+1}$ everywhere.

\noindent{\bf Mass conservation:}

For mass, in each cell we have
\begin{equation}
\rho^{n+1}_j=\rho^n_j+\frac{1}{\Delta x}
\int^{t_{n+1}}_{t_n} \left[F^r_{j-1/2,\rho}-F^l_{j+1/2,\rho}\right] \ud t,
\end{equation}
where $F^{r,l}_{j+1/2,\rho}$ are the mass fluxes. The total mass in
the control volume is $\sum_{j=K_1}^{K_2} \rho_j$, and
\begin{equation}\label{tmass}
\begin{array}{c}
\sum_{j=K_1}^{K_2}\rho^{n+1}_j=\sum_{j=K_1}^{K_2}\rho^n_j+\frac{1}{\Delta x}
\int^{t_{n+1}}_{t_n} \sum_{j=K_1}^{K_2}\left[F^r_{j-1/2,\rho}-F^l_{j+1/2,\rho}\right] \ud t.
\end{array}
\end{equation}
From (\ref{flux1-l}) and (\ref{flux1-r}), we have
\begin{equation}\label{masseq}
\begin{array}{l}
\quad F^l_{j+1/2,\rho}\\ \\
=\iint^{+\infty}_0 f_j(u)u \ud u \ud \xi
-\iint^{U_c}_0 f_j(u)u \ud u \ud \xi + \iint^0_{-\infty}
f_{j+1}(u)u \ud u \ud \xi\\ \\
=\iint^{+\infty}_{U_c} f_j(u)u \ud u \ud \xi
+ \iint^0_{-\infty}f_{j+1}(u)u \ud u \ud \xi\\ \\
=F^r_{j+1/2,\rho}.
\end{array}
\end{equation}
Therefore, from (\ref{tmass}) and (\ref{masseq}),
\begin{equation}
\begin{array}{c}
\sum_{j=K_1}^{K_2}\rho^{n+1}_j=\sum_{j=K_1}^{K_2}\rho^n_j+\frac{1}{\Delta x}
\int^{t_{n+1}}_{t_n} \left[F^r_{K_1-1/2,\rho}-F^l_{K_2+1/2,\rho}\right] \ud t,
\end{array}
\end{equation}
which gives the mass conservation in the computational domain.

\noindent{\bf Total energy conservation:}

The kinetic energy and thermal energy, i.e., $\rho E$, is updated by
\begin{equation}
(\rho E)^{n+1}_j=(\rho E)^n_j+\frac{1}{\Delta x}
\int^{t_{n+1}}_{t_n} \left[F^r_{j-1/2,\rho E}-F^l_{j+1/2,\rho E}\right] \ud t,
\end{equation}
where $F^{r,l}_{j+1/2,\rho E}$ are the fluxes of $\rho E$.
Because the external potential $\phi$ is independent of time, the potential energy,
i.e., $\rho \phi$ is updated by
\begin{equation}
\rho^{n+1}_j \phi_j=\rho^n_j \phi_j+\frac{1}{\Delta x}
\int^{t_{n+1}}_{t_n} \left[F^r_{j-1/2,\rho}\phi_j-F^l_{j+1/2,\rho}\phi_j\right] \ud t.
\end{equation}
With the definition of total energy $TE=\rho E+\rho \phi$, we get
\begin{equation}
\begin{array}{l}
TE^{n+1}_j=TE^n_j+\frac{1}{\Delta
x}\int^{t_{n+1}}_{t_n}\left[F^r_{j-1/2,\rho}\phi_j-F^l_{j+1/2,\rho}\phi_j \right.\\ \\
\qquad \qquad\left.+F^r_{j-1/2,\rho E}-F^l_{j+1/2,\rho E}\right] \ud t.
\end{array}
\end{equation}
The updating of the total energy in the control volume (i.e.
$\sum_{j=K_1}^{K_2} TE_j$) becomes
\begin{equation}\label{ttener}
\begin{array}{l}
\sum_{j=K_1}^{K_2} TE^{n+1}_j=\sum_{j=K_1}^{K_2} TE^n_j+\frac{1}{\Delta
x}\int^{t_{n+1}}_{t_n}\sum_{j=K_1}^{K_2}\left[F^r_{j-1/2,\rho}\phi_j
 \right.\\ \\
\qquad \qquad \qquad \quad \left.-F^l_{j+1/2,\rho}\phi_j+F^r_{j-1/2,\rho E}-F^l_{j+1/2,\rho E}\right] \ud t.
\end{array}
\end{equation}
According to (\ref{flux1-l}) and (\ref{flux1-r}), we get
\begin{equation}
\begin{array}{l}
F^l_{j+1/2,\rho E}=\iint^{+\infty}_0 f_j(u)\frac{1}{2}(u^3+u\xi) \ud u \ud \xi
+\iint^{U_c}_0 f_j(u)\frac{1}{2}(-u^3-u\xi) \ud u \ud \xi\\ \\
\qquad \qquad \quad+ \iint^0_{-\infty}f_{j+1}(u)\frac{1}{2}(u(u^2+U_c^2)+u\xi) \ud u \ud \xi,\\ \\
F^r_{j+1/2,\rho E}=\iint^{+\infty}_{U_c} f_j(u)\frac{1}{2}(u(u^2-U_c^2)+u\xi) \ud u \ud \xi\\ \\
\qquad \qquad \quad+ \iint^0_{-\infty}f_{j+1}(u)\frac{1}{2}(u^3+u\xi) \ud u \ud \xi.
\end{array}
\end{equation}
A direct calculation gives
\begin{equation}\label{enereq}
F^r_{j+1/2,\rho E}-F^l_{j+1/2,\rho E}=F^l_{j+1/2,\rho}(\phi_{j+1}-\phi_j)
=F^r_{j+1/2,\rho}(\phi_{j+1}-\phi_j).
\end{equation}
So, from (\ref{ttener}) and (\ref{enereq}), the total energy update
becomes
\begin{equation}\label{ttener2}
\begin{array}{l}
\sum_{j=K_1}^{K_2} TE^{n+1}_j=\sum_{j=K_1}^{K_2} TE^n_j+\frac{1}{\Delta
x}\int^{t_{n+1}}_{t_n}\left[F^r_{K_1-1/2,\rho}\phi_{K_1}
 \right.\\ \\
\qquad \qquad \qquad \quad \left.-F^l_{K_2+1/2,\rho}\phi_{K_2}+F^r_{K_1-1/2,\rho E}-F^l_{K_2+1/2,\rho E}\right] \ud t,
\end{array}
\end{equation}
which guarantees  the total energy conservation in the whole
computational domain. Based on the above proof, the SP-BGK and
SP-KFVS schemes are conservative methods. Therefore, the above two
schemes can give the correct shock location even with the external
gravitational forcing terms. This is a generalization of
Lax-Wendroff theorem to the
system with gravitational source term \cite{leveque-book}.\\

\noindent{\bf Lemma 3.2:} The density $\rho(x)$ in a hydrostatic
state  under the gravitational field $\phi(x)$ satisfies
\begin{equation}\label{eq:hydro-dens}
\rho(x)=C_1e^{-2\tilde{\lambda}\phi(x)},
\end{equation}
where $C_1$ and $\tilde{\lambda}$ are constants.

\noindent {\bf Proof} \quad For a hydrostatic solution under
the gravitational field $\phi(x)$, we have
\begin{equation}\label{eq:hydro}
p_x=-\rho \phi_x, T={\mbox{constant}}, U=0.
\end{equation}
Since $T={\mbox{constant}}$ and $\lambda=m/2kT$, we know
$\lambda=\tilde{\lambda}$, where $\tilde{\lambda}$ is also a
constant. Then from (\ref{eq:hydro}) and the ideal gas equation of
state
\[
p=\frac{1}{2\tilde{\lambda}} \rho,
\]
we have
\[
\frac{1}{2\tilde{\lambda}} \rho_x=-\rho \phi_x.
\]
Therefore, with a constant, $C_1$, the solution becomes
\[
\rho(x)=C_1e^{-2\tilde{\lambda}\phi(x)}.
\]
{\bf Remark:} without losing generality, in the following proofs, we
let $C_1=1$ for the hydrostatic solution. So, in the hydrostatic
case, the state has the form
\begin{equation}\label{eq:hydro-soln}
\rho=e^{-2\tilde{\lambda} \phi(x)},\,U=0,
\end{equation}
where $\tilde{\lambda}$ is a constant. Numerically, if we let the
potential $\phi(x)$ be a constant, $\phi_j$, in the $jth$ cell, then
\begin{equation}\label{eq:hydro-soln-numer}
\rho_{j+1}=\rho_j e^{-2\tilde{\lambda} (\phi_{j+1}-\phi_j)},\,
U_j=0,
\end{equation}\\
where $\rho_j$ and $U_j$ are cell average quantities in that cell.\\

\noindent{\bf Lemma 3.3:} For the two equilibrium states
$W^l_{j+1/2} =(\rho^l_{j+1/2},\,(\rho U)^l_{j+1/2}, \,(\rho
E)^l_{j+1/2})$ and $W^r_{j+1/2}=(\rho^r_{j+1/2},\,(\rho
U)^r_{j+1/2}, \,(\rho E)^r_{j+1/2})$, they have the following
properties when the initial flow is in a hydrostatic state.

1. Both velocities are equal to zero, i.e.,
\begin{equation}\label{eq:velo-zero}
U^l_{j+1/2}=U^r_{j+1/2}=0.
\end{equation}

2. They have the same temperature at both sides of all cell
interfaces, i.e.
\begin{equation}\label{eq:tem-equal}
\lambda^l_{j+1/2}=\lambda^r_{j+1/2}=\tilde{\lambda},
\end{equation}
where $\lambda$ satisfies
\begin{equation}\label{eq:lambda-mac}
\rho E- \frac{1}{2} \rho U^2= \rho \frac{K+1}{4 \lambda},
\end{equation}
macroscopically with $K=(3-\gamma)/(\gamma-1)$ in 1-D, and
$\tilde{\lambda}$ has the constant value $\lambda$ of the
hydrostatic solution.

3. The densities at the same cell interface satisfy
\begin{equation}\label{eq:dens-hydroeq}
\rho^r_{j+1/2}=\rho^l_{j+1/2}e^{-2\tilde{\lambda}(\phi_{j+1}-\phi_j)}
\end{equation}

4. In the same cell,
\begin{equation}\label{eq:dens-equal}
\rho^l_{j+1/2}=\rho^r_{j-1/2}
\end{equation}

\noindent {\bf Proof} \quad As the definition, $W^l_{j+1/2}$ and
$W^r_{j+1/2}$ are determined  by (\ref{var1-l}) and (\ref{var1-r})
or (\ref{var2-l}) and (\ref{var2-r}) for $\phi_j<\phi_{j+1}$ or
$\phi_j>\phi_{j+1}$ when $f_j(u)=g_j(u)$, where $g_j(u)$ is a
Maxwellian corresponding to the cell average conservative variables,
$(\rho_j,\,(\rho U)_j, \,(\rho E)_j)$. Here, we only prove the case
for $\phi_j<\phi_{j+1}$. The other case can be proved similarly.
From direct calculation, we can get
\begin{equation}\label{eq:equi-densl}
\rho^l_{j+1/2} = \frac{\rho_j}{2} + \rho_j(\frac{\tilde{\lambda}}{\pi})^{\frac{1}{2}} \int_{-U_c}^{0} e^{-\tilde{\lambda} u^2}\mathrm{d}u
- \rho_{j+1} (\frac{\tilde{\lambda}}{\pi})^{\frac{1}{2}} U_c + \rho_{j+1} \tilde{\lambda} (\frac{\tilde{\lambda}}{\pi})^{\frac{1}{2}}\int_{0}^{+\infty} e^{-\tilde{\lambda} t}\sqrt{t+ U_c^2}\mathrm{d}t,
\end{equation}

\begin{equation}\label{eq:equi-densr}
\rho^r_{j+1/2} =  \rho_j  \tilde{\lambda} (\frac{\tilde{\lambda}}{\pi})^{\frac{1}{2}} \int_{U_c^2}^{+\infty} e^{-\tilde{\lambda} t}\sqrt{t - U_c^2}\mathrm{d}t
+\frac{\rho_{j+1}}{2},
\end{equation}

\begin{equation}\label{eq:equi-mom}
(\rho U)^l_{j+1/2}=(\rho U)^r_{j+1/2}=0,
\end{equation}

\begin{equation}\label{eq:equi-enerl}
\begin{array}{l}(\rho E)^l_{j+1/2} = \frac{K}{4\tilde{\lambda}} \rho^l_{j+1/2} + \frac{\rho_{j}}{8 \tilde{\lambda}} - \frac{\rho_{j}}{4\tilde{\lambda}}\sqrt{\frac{\tilde{\lambda}}{\pi}}e^{-\tilde{\lambda} U_c^2 } U_c + \frac{\rho_{j}}{4\tilde{\lambda}}\sqrt{\frac{\tilde{\lambda}}{\pi}} \int_{-U_c}^{0}e^{-\tilde{\lambda} u^2 } \mathrm{d}u \\ \\
\qquad \qquad \qquad+\frac{\rho_{j+1}}{4}\sqrt{\frac{\tilde{\lambda}}{\pi}} \int_{0}^{+\infty}e^{-\tilde{\lambda} t } \sqrt{t + U_c^2} \mathrm{d}t,
\end{array}
\end{equation}

and

\begin{equation}\label{eq:equi-enerr}
(\rho E)^r_{j+1/2} = \frac{K}{4 \tilde{\lambda}} \rho^r_{j+1/2}+ \frac{\rho_{j}}{4}\sqrt{\frac{\tilde{\lambda}}{\pi}} \int_{U_c^2}^{+\infty}e^{-\tilde{\lambda} t } \sqrt{t - U_c^2} \mathrm{d}t
 + \frac{\rho_{j+1}}{8 \tilde{\lambda}},
\end{equation}
where $U_c=\sqrt{2(\phi_{j+1}-\phi_j)}$.\\

1. From (\ref{eq:equi-densl}) and (\ref{eq:equi-densr}), we can easily see that
$\rho^l_{j+1/2}>0$ and $\rho^r_{j+1/2}>0$ when $\rho_j>0$ and $\rho_{j+1}>0$. Since
$U=\rho U/\rho$, from (\ref{eq:equi-mom}), we know that
\[
U^l_{j+1/2}=U^r_{j+1/2}=0.
\]

2. From (\ref{eq:equi-densl}),
\begin{equation}\label{eq:lem2-1}
\begin{array}{l}\rho^l_{j+1/2}\,\frac{K+1}{4 \lambda^l_{j+1/2}}=\frac{K}{4 \lambda^l_{j+1/2}}\rho^l_{j+1/2}  +
\frac{\rho_j}{8\lambda^l_{j+1/2}} + \frac{\rho_{j+1}}{4\lambda^l_{j+1/2}}\sqrt{\frac{\tilde{\lambda}}{\pi}} U_c +
\frac{\rho_{j}}{4\lambda^l_{j+1/2}}\sqrt{\frac{\tilde{\lambda}}{\pi}} \int_{-U_c}^{0}e^{-\tilde{\lambda}u^2 } \mathrm{d}u  \\ \\
\qquad \qquad \qquad \quad +\frac{\rho_{j+1}}{4\lambda^l_{j+1/2}} \tilde{\lambda} \sqrt{\frac{\tilde{\lambda}}{\pi}} \int_{0}^{+\infty}e^{-\tilde{\lambda} t } \sqrt{t + U_c^2} \mathrm{d} t.
\end{array}
\end{equation}
Since $(\rho E)^l_{j+1/2}- \frac{1}{2} \rho^l_{j+1/2} (U^l_{j+1/2})^2
= \rho^l_{j+1/2}\, \frac{K+1}{4 \lambda^l_{j+1/2}}$ and $U^l_{j+1/2}=0$, we have
\begin{equation}\label{eq:lem2-2}
(\rho E)^l_{j+1/2}
-\rho^l_{j+1/2} \frac{K+1}{4 \lambda^l_{j+1/2}}=0.
\end{equation}
Therefore, substitute  (\ref{eq:hydro-soln-numer}),
(\ref{eq:equi-enerl}) and (\ref{eq:lem2-1}) into (\ref{eq:lem2-2}),
we get
\begin{equation}\label{eq:lem2-3}
\begin{array}{l}(\lambda^l_{j+1/2} - \tilde{\lambda})\left\{\frac{1}{\lambda^l_{j+1/2}\tilde{\lambda}}
(\frac{K}{4}\rho^l_{j+1/2} +\frac{\rho_j}{8} - \frac{\rho_{j}}{4}\sqrt{\frac{\tilde{\lambda}}{\pi}}e^{-\tilde{\lambda} U_c^2 } U_c \right.\\ \\
\left.+ \frac{\rho_{j}}{4}\sqrt{\frac{\tilde{\lambda}}{\pi}} \int_{-U_c}^{0}e^{-\tilde{\lambda} u^2 }\mathrm{d}u )+
\frac{1}{\lambda^l_{j+1/2}} \frac{\rho_{j+1}}{4}\sqrt{\frac{\tilde{\lambda}}{\pi}} \int_{0}^{+\infty}e^{-\tilde{\lambda} t } \sqrt{t + U_c^2} \mathrm{d}t \right\} = 0.
\end{array}
\end{equation}
Because $e^{-\tilde{\lambda} u^2 }$ is a monotonic increasing function on $[-U_c,0]$, so
\begin{equation}\label{eq:lem2-4}
\frac{\rho_{j}}{4}\sqrt{\frac{\lambda}{\pi}} \int_{-U_c}^{0}e^{-\lambda u^2 }\mathrm{d}u - \frac{\rho_{j}}{4}\sqrt{\frac{\lambda}{\pi}}e^{-\lambda U_c^2 } U_c > 0.
\end{equation}
Then we know the summation in the brace $\{... \}$ of
(\ref{eq:lem2-3}) is strictly larger than zero. Therefore,
\[
\lambda^l_{j+1/2}=\tilde{\lambda}.
\]
has to be satisfied.

Similarly, we can have
\[
(\lambda^r_{j+1/2} - \tilde{\lambda})\{\frac{1}{\lambda^r_{j+1/2}\tilde{\lambda}}
(\frac{K}{4}\rho^r_{j+1/2} + \frac{\rho_{j+1}}{8} ) + \frac{1}{\lambda^r_{j+1/2}}
\frac{\rho_{j}}{4}\sqrt{\frac{\tilde{\lambda}}{\pi}} \int_{U_c^2}^{+\infty}e^{-\tilde{\lambda} t } \sqrt{t - U_c^2} \mathrm{d}t \} = 0.
\]
Again, the summation in the brace $\{ ... \}$ is strictly larger
than zero. So,
\[
\lambda^r_{j+1/2}=\tilde{\lambda}.
\]

3. It is easy to prove that
\begin{equation}\label{eq:lem2-5}
\int_{-U_c}^{0} e^{-\tilde{\lambda} u^2}\mathrm{d}u =e^{-\lambda U_c^2 } U_c + 2 \tilde{\lambda} \int_{-U_c}^{0} e^{-\tilde{\lambda} u^2} u^2 \mathrm{d}u,
\end{equation}
and
\begin{equation}\label{eq:lem2-6}
2 \int_{-U_c}^{0}e^{-\tilde{\lambda} u^2} u^2 \mathrm{d}u= \int_{0}^{U_c^2}e^{-\tilde{\lambda} x} \sqrt{x} \mathrm{d}x .
\end{equation}
So,
\qquad $\rho^r_{j+1/2}=\rho^l_{j+1/2}e^{-2\tilde{\lambda}(\phi_{j+1}-\phi_j)},$\\ \\
$\xLongleftrightarrow{(\ref{eq:equi-densl}), (\ref{eq:equi-densr})}$
$\tilde{\lambda} \int_{U_c^2}^{+\infty}e^{-\tilde{\lambda}t } \sqrt{t - U_c^2} \mathrm{d}t  = \int_{-U_c}^{0} e^{-\tilde{\lambda} (u^2 + U_c^2)}\mathrm{d}u  - U_c e^{-2 \tilde{\lambda} U_c^2}+\tilde{\lambda} e^{-2\tilde{\lambda}U_c^2}\int_{0}^{+\infty}e^{-\tilde{\lambda} t } \sqrt{t + U_c^2} \mathrm{d}t,$
$\xLongleftrightarrow{(\ref{eq:lem2-5})}$
$\int_{U_c^2}^{+\infty}e^{-\tilde{\lambda} t } \sqrt{t - U_c^2} \mathrm{d}t = 2 e^{-\tilde{\lambda} U_c^2}\int_{-U_c}^{0} e^{-\tilde{\lambda} u^2} u^2 \mathrm{d}u + e^{-2\tilde{\lambda} U_c^2}\int_{0}^{+\infty}e^{-\tilde{\lambda} t } \sqrt{t + U_c^2} \mathrm{d}t, $\\ \\
$\xLongleftrightarrow{left: x=t-U_c^2; right: x=t+U_c^2}$
$\int_{0}^{+\infty}e^{-\tilde{\lambda} x} \sqrt{x} \mathrm{d}x = 2 \int_{-U_c}^{0}e^{-\tilde{\lambda} u^2} u^2 \mathrm{d}u + \int_{U_c^2}^{+\infty}e^{-\tilde{\lambda} x} \sqrt{x} \mathrm{d}x$\\ \\
$\Longleftrightarrow \int_{0}^{U_c^2}e^{-\tilde{\lambda} x} \sqrt{x} \mathrm{d}x = 2 \int_{-u_c}^{0}e^{-\tilde{\lambda} u^2} u^2 \mathrm{d}u. $\\

Therefore, from (\ref{eq:lem2-6}), we can conclude that
\[\rho^r_{j+1/2}=\rho^l_{j+1/2}e^{-2\tilde{\lambda}(\phi_{j+1}-\phi_j)}.\]

4.
\qquad $\rho^l_{j+1/2}=\rho^r_{j-1/2},$\\ \\
$\xLongleftrightarrow{(\ref{eq:equi-densl}), (\ref{eq:equi-densr})}$
\quad$\frac{\rho_j}{2} + \rho_j(\frac{\tilde{\lambda}}{\pi})^{\frac{1}{2}} \int_{-U_c}^{0} e^{-\tilde{\lambda} u^2}\mathrm{d}u
- \rho_{j+1} (\frac{\tilde{\lambda}}{\pi})^{\frac{1}{2}} U_c + \rho_{j+1} \tilde{\lambda} (\frac{\tilde{\lambda}}{\pi})^{\frac{1}{2}}\int_{0}^{+\infty} e^{-\tilde{\lambda} t}\sqrt{t+ U_c^2}\mathrm{d}t$\\

\qquad $=
\rho_{j-1} \tilde{\lambda}(\frac{\tilde{\lambda}}{\pi})^{\frac{1}{2}}  \int_{U_c^2}^{+\infty} e^{-\tilde{\lambda} t}\sqrt{t - U_c^2}\mathrm{d}t
+\frac{\rho_j}{2},$\\ \\
$\xLongleftrightarrow{(\ref{eq:hydro-soln-numer})} \rho_j \int_{-U_c}^{0} e^{-\tilde{\lambda} u^2}\mathrm{d}u
- \rho_j e^{-\tilde{\lambda} U_c^2}  U_c + \rho_j \tilde{\lambda} \int_{U_c^2}^{+\infty} e^{-\tilde{\lambda} x}\sqrt{x}\mathrm{d}x=
\rho_j \tilde{\lambda} \int_{0}^{+\infty} e^{-\tilde{\lambda} x}\sqrt{x}\mathrm{d}x,$\\ \\
$\xLongleftrightarrow{(\ref{eq:lem2-5})}$
$2\tilde{\lambda}\rho_j \int_{-U_c}^{0} e^{-\tilde{\lambda} u^2} u^2\mathrm{d}u
+ \rho_j \tilde{\lambda} \int_{U_c^2}^{+\infty} e^{-\tilde{\lambda} x}\sqrt{x}\mathrm{d}x=
\rho_j \tilde{\lambda} \int_{0}^{+\infty} e^{-\tilde{\lambda} x}\sqrt{x}\mathrm{d}x.$\\ \\
From (\ref{eq:lem2-6}), we know that the last equality holds. Therefore,
\[\rho^l_{j+1/2}=\rho^r_{j-1/2}.\]

\noindent{\bf Remark:} the above lemma, especially part 2,
illustrates that starting from a hydrostatic state with the same
temperature, the constructed equilibrium states at both sides of a
cell interface have the equal temperature as well. In order
words, in the hydrostatic case, the particle interaction with the
potential barrier and the particle collisions among themselves never
alter the equilibrium temperature both sides of a cell
interface. This is consistent with the second law of thermodynamics.
Otherwise, the temperature differences generated by the particle
collisions could be used drive an engine and a pure work could have
been extracted
from an initially isothermal system. This violates the 2nd-law of thermodynamics. \\

\noindent{\bf Theorem 3.4:} For a well-balanced kinetic scheme, the
equilibrium distribution function must be an "Exact Maxwellian".

\noindent {\bf Proof} \quad In order to keep the hydrostatic solution (\ref{eq:hydro-soln-numer})
the numerical
mass flux at both sides of a cell interface must be zero.

Without losing generality, we only consider the case for
$\phi_{j+1}>\phi_j$. Since the gas must be isotropic, we can assume
the equilibrium distribution function is $\rho(x) G(u^2)$ and define
$a=\sqrt{2(\phi_{j+1}-\phi_j)}$, then we require
\begin{equation}
F^r_{j+1/2,\rho}=\int^{+\infty}_a \rho_j G(u^2)u \ud u
+\int^0_{-\infty} \rho_{j+1} G(u^2)u \ud u =0,
\end{equation}
where $F^r_{j+1/2,\rho}$ is the mass flux at the right side of the interface.
Because of (\ref{eq:hydro-soln-numer}), we have
\begin{equation}\label{1-mf}
\frac{1}{2} \int^{+\infty}_{a^2} G(x) \ud x + e^{-\lambda a^2}
\int^0_{-\infty}G(u^2)u \ud u =0.
\end{equation}
Take the derivative of (\ref{1-mf}) with $a^2$, we get
\begin{equation}\label{1-mf1}
-\frac{1}{2} G(a^2)-\lambda e^{-\lambda a^2}\int^0_{-\infty}G(u^2)u
\ud u =0.
\end{equation}
It is obvious from (\ref{1-mf1}) that
\begin{equation}
G(a^2)\sim e^{-\lambda a^2},
\end{equation}
which means that the equilibrium distribution function is an exact Maxewellian distribution.\\

\noindent{\bf Theorem 3.5:} Both the 1st-order SP-KFVS and SP-BGK
schemes are well-balanced schemes.

\noindent{\bf Proof} \quad In order to prove a scheme to be a
well-balanced one, we only need to verify that the scheme can keep
the hydrostatic solution (\ref{eq:hydro-soln}) forever. Numerically,
the initial condition for this case  is given by
(\ref{eq:hydro-soln-numer}) in the $jth$ cell. At the next time
step, the above solution must be kept by the well-balanced numerical
scheme, i.e., $W^{n+1}_j=W^n_j$. From (\ref{fv}), we must have
\begin{equation}\label{fluxeq}
F^r_{j-1/2}=F^l_{j+1/2}.
\end{equation}
Therefore, to complete the proof, we have to show that mass fluxes
($F^{r,l}_{j+1/2,\rho}$), momentum fluxes ($F^{r,l}_{j+1/2,\rho U}$)
and energy fluxes ($F^{r,l}_{j+1/2,\rho E}$) satisfy the condition
(\ref{fluxeq}) respectively. \\

{\bf The $1st$-order SP-KFVS scheme:} the original distribution
function at the cell interface is
\begin{equation}
f(x_{j+1/2},t,u,\xi)=\left\{\begin{array}{ll}g_j(u), & u \ge 0,\\ \\
g_{j+1}(u), & u<0, \end{array}\right.
\end{equation}
where $g_j(u)$ is the Maxwellian corresponding to $(\rho_j,\,(\rho
U)_j, \,(\rho E)_j)$. The proof is only a direct calculation of the
fluxes at the interface using (\ref{flux1-l}) and (\ref{flux1-r}) or
(\ref{flux2-l}) and (\ref{flux2-r}) in two different cases for
$\phi_j<\phi_{j+1}$ or $\phi_j>\phi_{j+1}$. Also the initial
hydrostatic condition (\ref{eq:hydro-soln-numer}) will be used.
The results are  the followings.\\
 a.\quad For mass flux,
\begin{equation}
F^l_{j+1/2,\rho}=F^r_{j+1/2,\rho}=0.
\end{equation}
b.\quad For momentum flux,
\begin{equation}
F^l_{j+1/2,\rho U}=F^r_{j-1/2,\rho U}=\frac{\rho_j}{2\lambda}.
\end{equation}
c.\quad For energy flux,
\begin{equation}
F^l_{j+1/2,\rho E}=F^r_{j+1/2,\rho E}=0.
\end{equation}
Hence, the first order $1st$ order SP-KFVS scheme is a well-balanced one.\\

{\bf The 1st order SP-BGK scheme:} the original distribution
function is
\begin{equation}
f(x_{j+1/2},t,u,\xi)=\left\{\begin{array}{ll}(1-\epsilon) g_j(u)+\epsilon g_{j+1/2}^l(u), & u \ge 0,\\ \\
(1-\epsilon)g_{j+1}(u)+\epsilon g_{j+1/2}^r(u), & u<0, \end{array}\right.
\end{equation}
where $\epsilon$ is a constant between $0$ and $1$, $g_j(u)$ is the
same as in the proof for the 1st order SP-KFVS scheme, $g_{j+1/2}^l$
and $g_{j+1/2}^r$ are two equilibrium states corresponding to
$W_{j+1/2}^l$ and $W_{j+1/2}^r$ respectively. Here, $W_{j+1/2}^l$
and $W_{j+1/2}^r$ are the macroscopic variables calculated by
(\ref{var1-l}) and (\ref{var1-r}) or (\ref{var2-l}) and
(\ref{var2-r}) when
\[
f_j(u)=g_j(u)\,\,\textrm{and}\,\,f_{j+1}(u)=g_{j+1}(u).
\]
So, the fluxes are the linear combination of two kinds of fluxes
$F_1$ and $F_2$ calculated by
\[
f_1=\left\{\begin{array}{ll}g_j(u), & u \ge 0,\\ \\
g_{j+1}(u), & u<0, \end{array}\right.
\,\,\textrm{and}\,\,
f_2=\left\{\begin{array}{ll}g_{j+1/2}^l(u), & u \ge 0,\\ \\
g_{j+1/2}^r(u), & u<0, \end{array}\right.
\]
respectively.

From the above proof for the 1st order SP-KFVS scheme, we know that
the first kind fluxes $F_1$ can satisfy (\ref{fluxeq}) itself.
Therefore, we only need to prove that $F_2$ can satisfy
(\ref{fluxeq}), too. Note that in the proof for the 1st order
SP-KFVS scheme, the hydrostatic initial condition is the key. But
from the Lemma 3.3, we can see that the equilibrium states also
satisfy the hydrostatic initial condition. So, similarly, we get the
following results for the fluxes corresponding to $f_2$ from a
direct calculation by using (\ref{flux1-l}) and (\ref{flux1-r}) or
(\ref{flux2-l}) and (\ref{flux2-r}) in two
different cases for $\phi_j<\phi_{j+1}$ or $\phi_j>\phi_{j+1}$.\\
a.\quad For mass flux,
\begin{equation}
F^l_{j+1/2,\rho}=F^r_{j+1/2,\rho}=0.
\end{equation}
b.\quad For momentum flux,
\begin{equation}
\begin{array}{l}
F^l_{j+1/2,\rho U}=\frac{\rho^l_{j+1/2}}{2\lambda^l_{j+1/2}},\\ \\
F^r_{j-1/2,\rho U}=\frac{\rho^r_{j-1/2}}{2\lambda^r_{j-1/2}}.
\end{array}
\end{equation}
Based on Eq.(\ref{eq:tem-equal}) and (\ref{eq:dens-equal}),
\begin{equation}
 F^l_{j+1/2,\rho U}=F^r_{j-1/2,\rho U}.
\end{equation}
c.\quad For energy flux,
\begin{equation}
F^l_{j+1/2,\rho E}=F^r_{j+1/2,\rho E}=0.
\end{equation}

From all the above proofs, we can conclude that both the 1st-order
SP-KFVS and SP-BGK schemes can keep the initial hydrostatic solution
forever. Therefore, they are well-balanced schemes.\\

\noindent{\bf Remark:} The 2nd order SP-KFVS and SP-BGK schemes are
well-balanced schemes.

We use $(U,\lambda , \rho e^{2\lambda\phi})$ to do the
reconstruction.  All the three variables are constants when the
solution is in a hydrostatic state. So, the slopes are all zeros
after using the MUSCL-type limiter. In other words, the 2nd-order
schemes go back to the 1st-order method when the solution is in
hydrostatic state, which can be kept forever. Therefore, the
2nd-order schemes are also well-balanced schemes.

\section{Numerical examples}\label{sec:tests}

In this section, we will present numerical results of four 1-D
examples by using $1^{st}$ and $2^{nd}$ order SP-KFVS and SP-BGK
schemes, and also a 2-D example using a $2^{nd}$-order SP-BGK
scheme. Each of the examples is very sensitive to the accuracy of
the scheme. Some of the tests run for millions of numerical steps.
If the scheme is not a  well-balanced one, the accumulation of any
small numerical error would become significant for such a long time
integration \cite{tian}.

\subsection{Shock tube under gravitational field}\label{shock}

\begin{figure}
\begin{center}
\includegraphics[scale=0.35,bb = 100 40 700 550, clip=true]{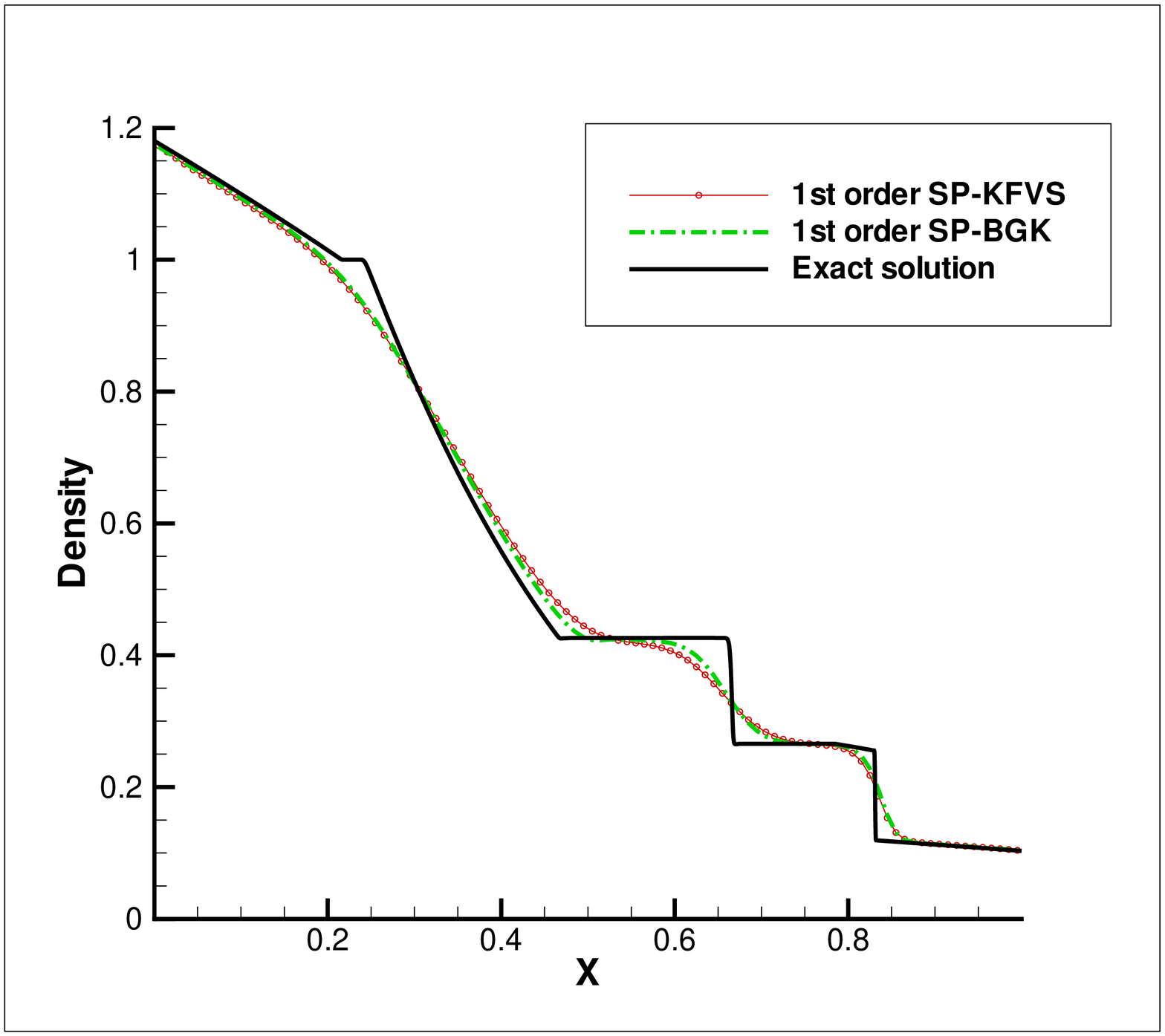}
\includegraphics[scale=0.35,bb = 100 40 700 550, clip=true]{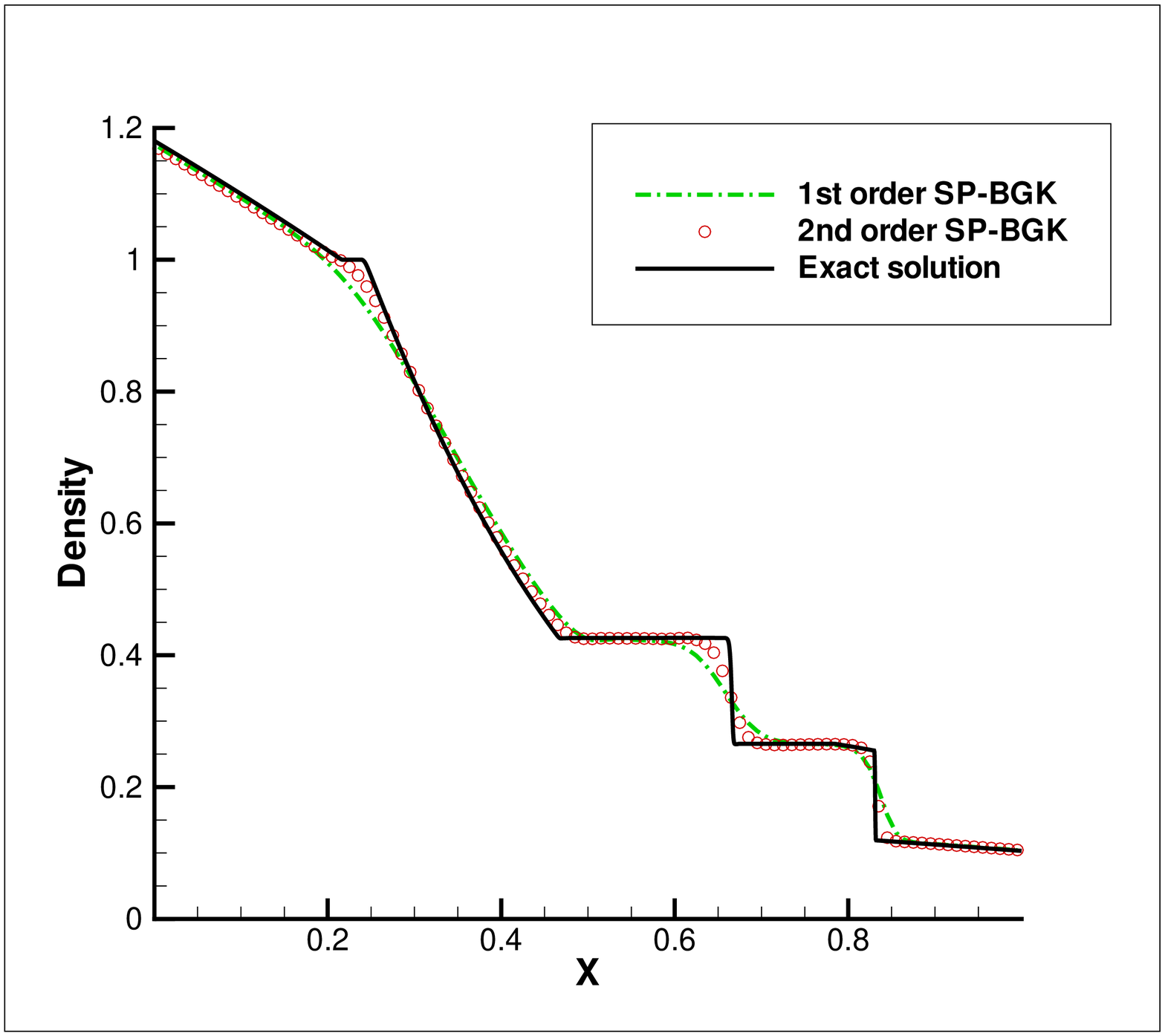}
\caption{\small Density distributions for the shock tube problem under gravitational field.}\label{shock-dens}
\end{center}
\end{figure}

\begin{figure}
\begin{center}
\includegraphics[scale=0.35,bb = 100 40 700 550, clip=true]{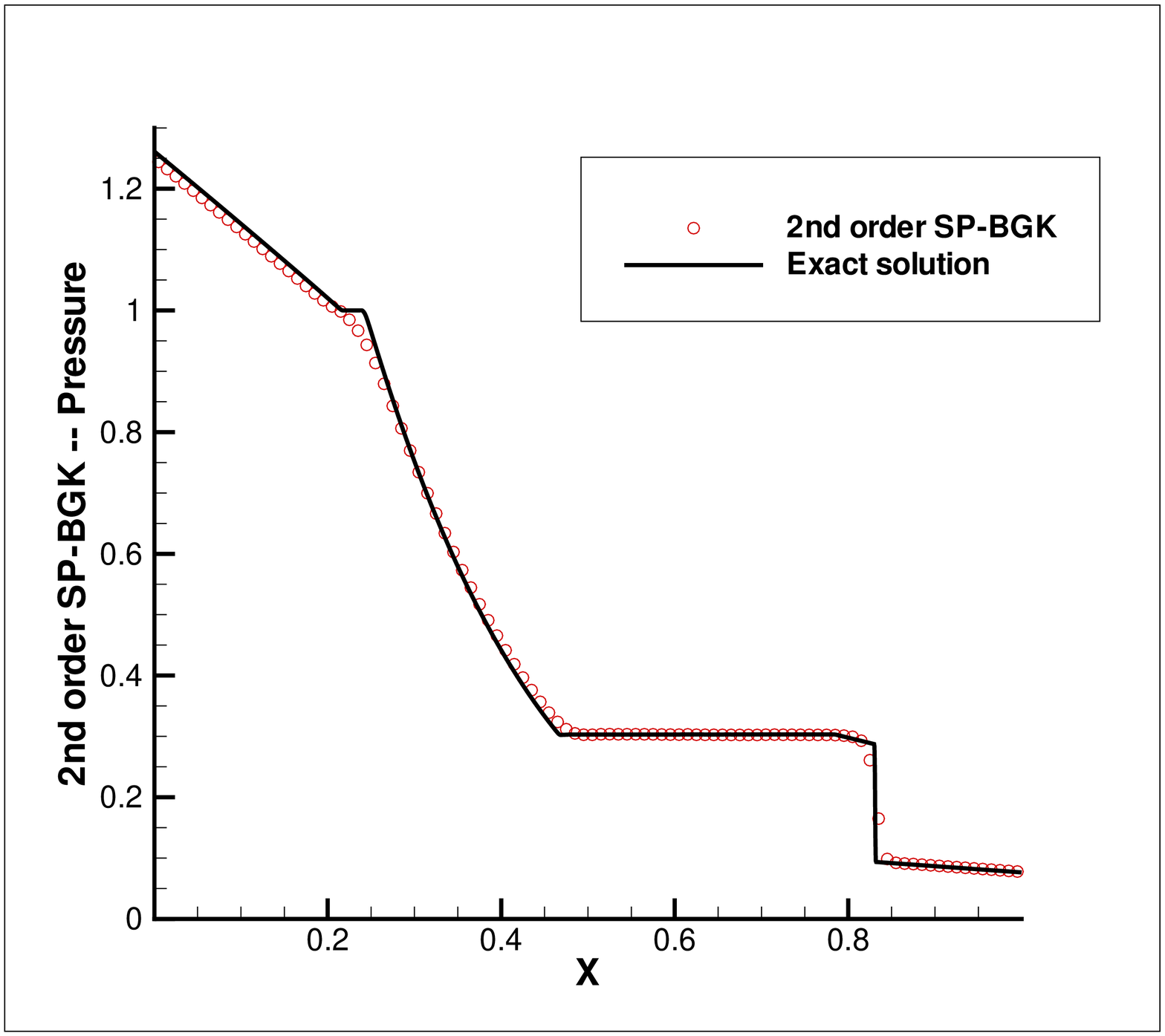}
\includegraphics[scale=0.35,bb = 100 40 718 590, clip=true]{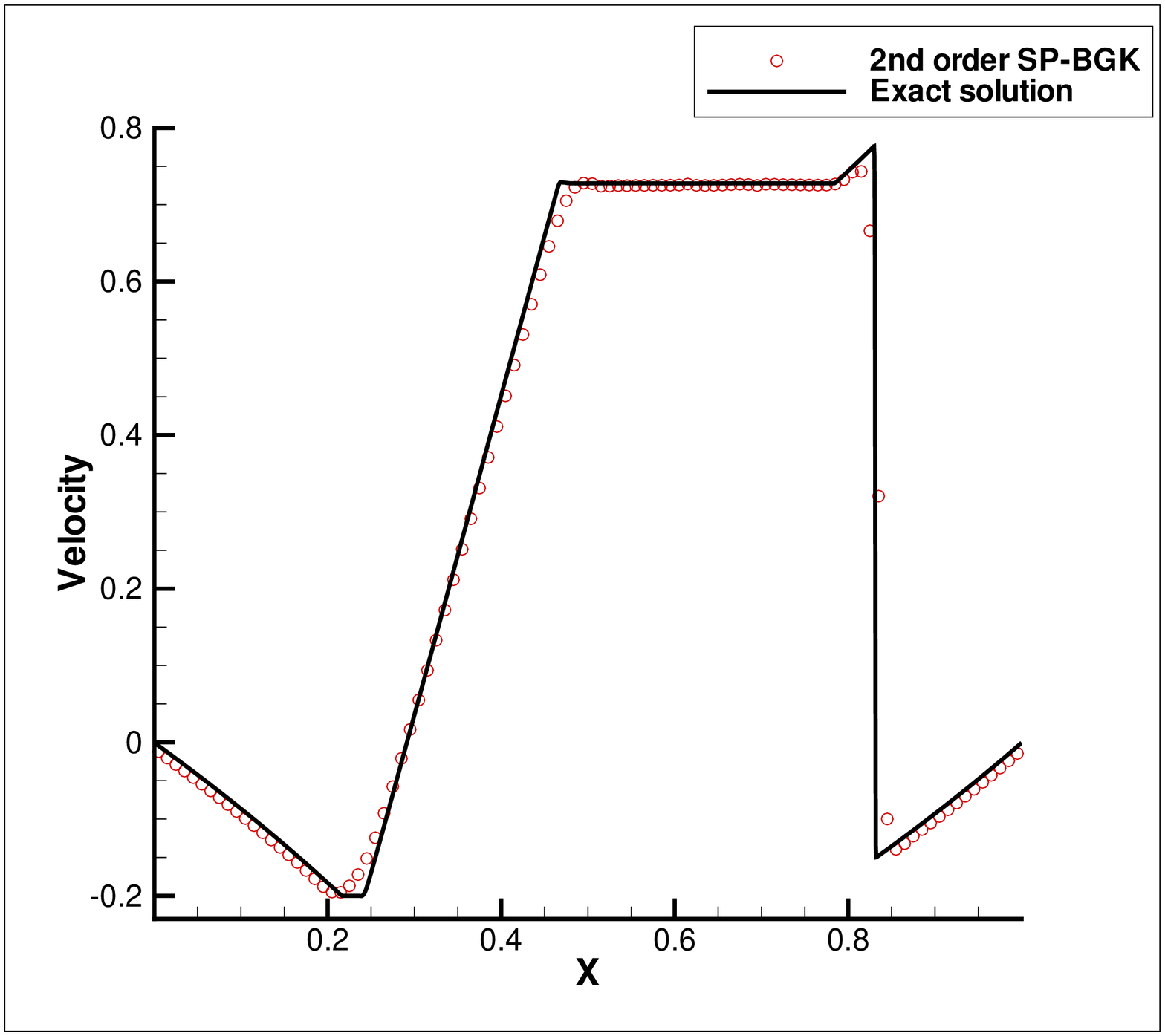}
\caption{\small Pressure and velocity distributions for the shock tube problem under gravitational field.}\label{shock-pres}
\end{center}
\end{figure}

This case is the standard Sod test under gravitational field. The
computational domain is $x\in [0,1]$ which is divided into $100$
cells. Reflection boundary condition is used on both ends. The
initial condition is
\[
\rho=1.0,U=0.0,p=1.0\textrm{ for }x\le 0.5,
\]
and
\[
\rho=0.125,U=0.0,p=0.1\textrm{ for }x> 0.5.
\]
The gravitational force $G$ takes a value $G=-1.0$ in the x-direction. So the potential jump at each cell interface becomes
\[\Delta \phi=-G\Delta x=0.01.\]
The computational results at $t=0.2$ are presented in
fig.~\ref{shock-dens}, ~\ref{shock-pres} for
the density, pressure and velocity from the $1^{st}$-order SP-KFVS,
$1^{st}$ and $2^{nd}$-order SP-BGK schemes. From these figures, we
can find that SP-KFVS scheme has larger numerical dissipation than
that in SP-BGK scheme, and $1^{st}$-order scheme is more dissipative
than $2^{nd}$-order one. The results calculated by the $2^{nd}$
order SP-BGK scheme fits the exact solution very well. Due to the
gravitational force, the density distribution inside the tube is
pulled back in the negative x-direction. In some region, the flow
velocity even becomes negative.

\subsection{Isolated gravitational system with adiabatic wall}

\begin{figure}
\begin{center}
\includegraphics[scale=0.35,bb = 100 40 700 550, clip=true]{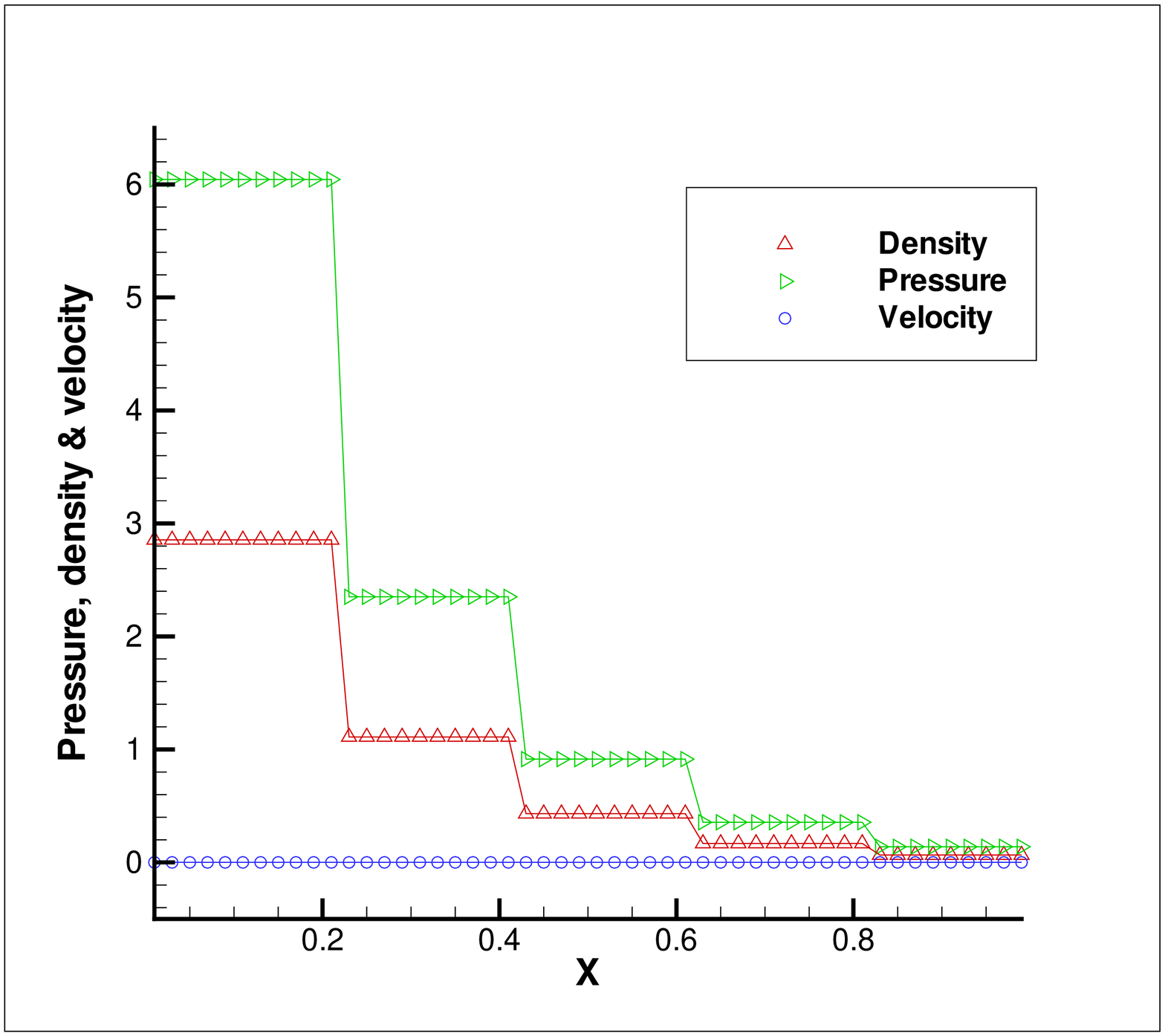}
\includegraphics[scale=0.35,bb = 80 40 700 550, clip=true]{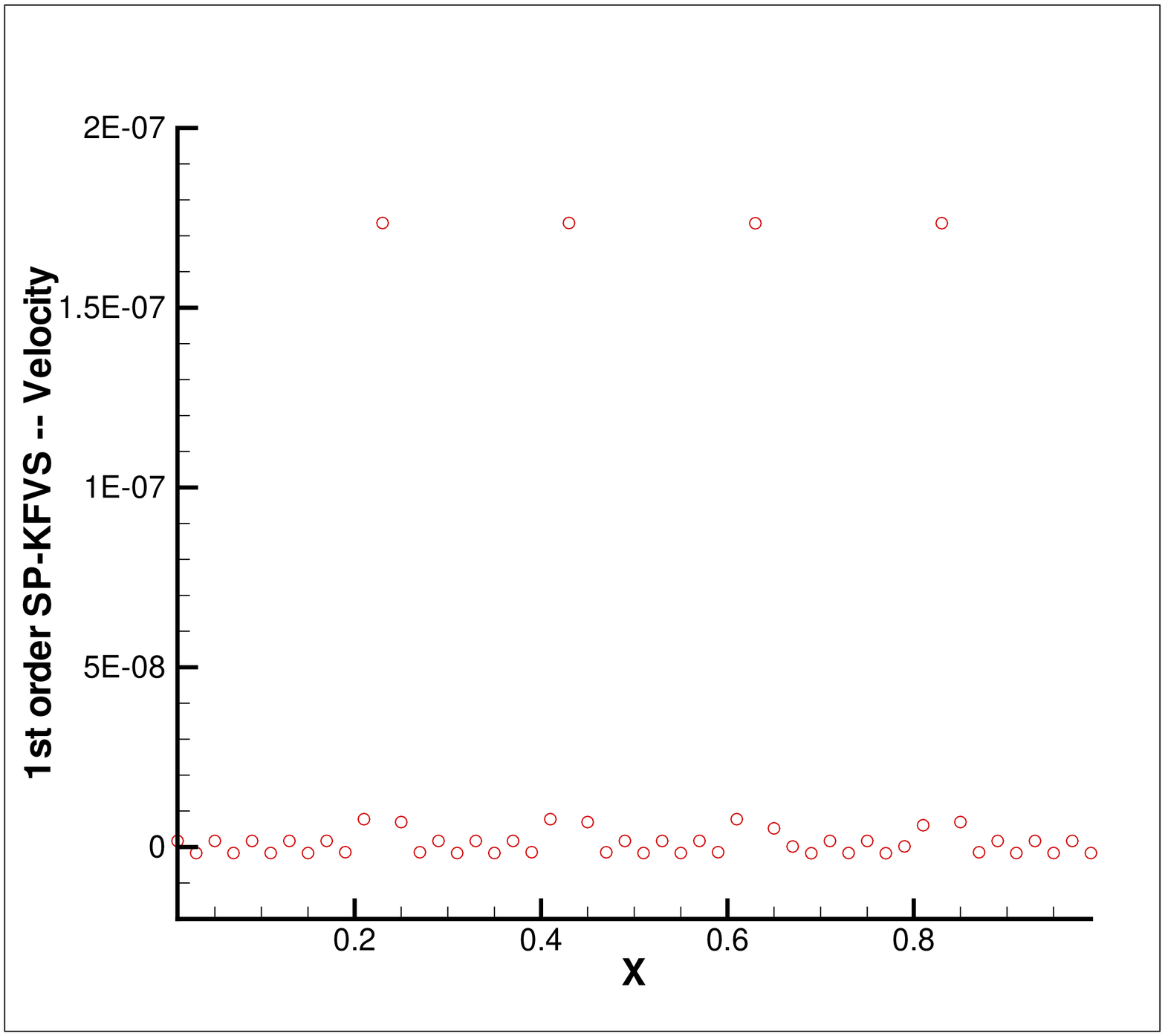}
\includegraphics[scale=0.35,bb = 90 40 700 550, clip=true]{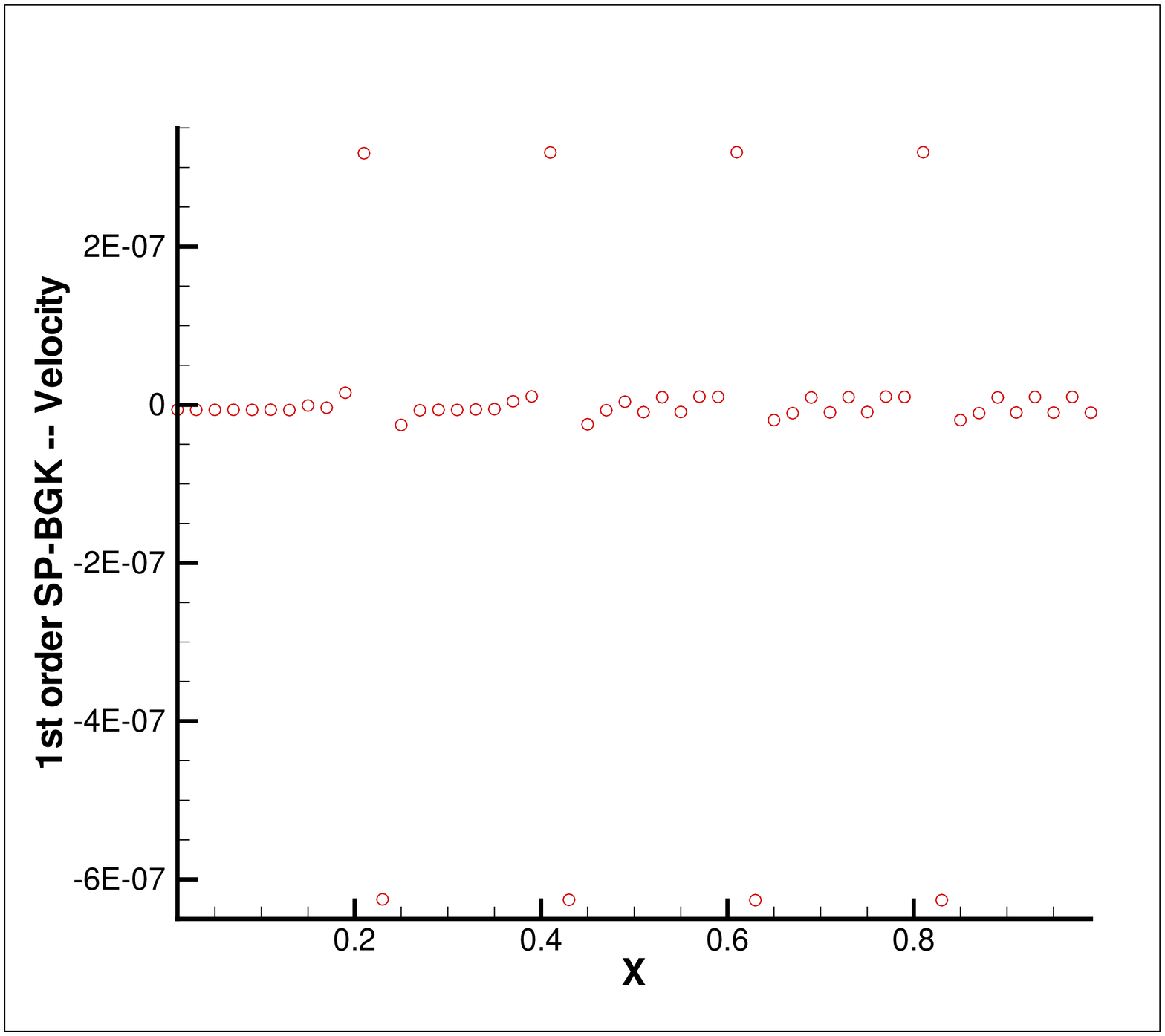}
\includegraphics[scale=0.35,bb = 80 40 700 550, clip=true]{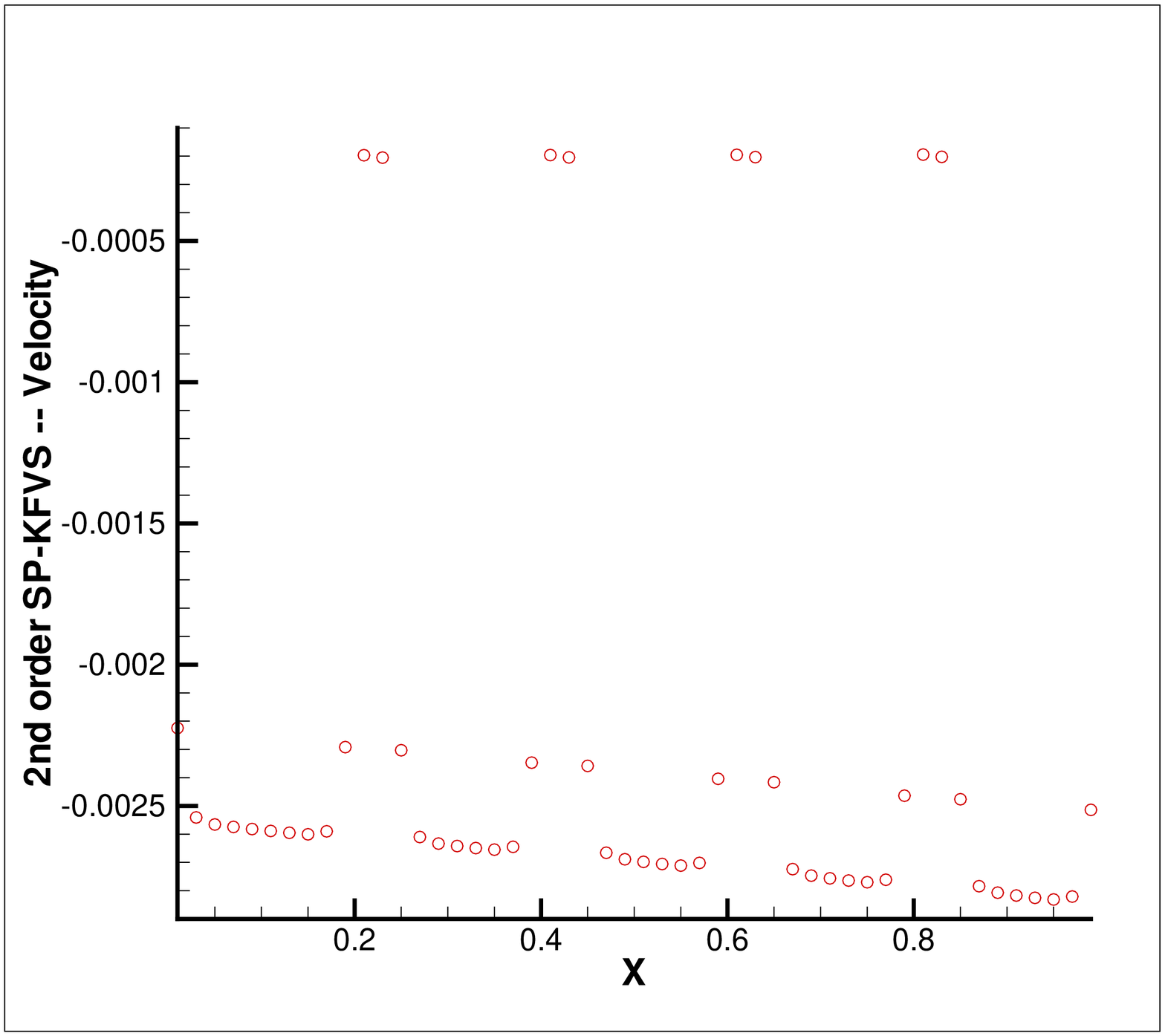}
\caption{\small The first figure shows the Density, pressure and velocity distributions calculated by $2^{nd}$ order SP-BGK for isolated gravitational system with adiabatic wall. Other figures are velocity distributions in this test case.}\label{iso-xvelo}
\end{center}
\end{figure}

The second test case is also on a computational domain $x \in [0,1]$
with $50$ cells. There are limited number of gravitational potential
jumps at locations $x=0.21, 0.41, 0.61$ and $0.81$ with a large
value
\[\Delta \Phi =2.0.\]
The initial flow distributions  inside the domain has  constant
values of
\[\rho =1.0, \rho U =0.0,\textrm{ and }\rho E =2.5.\]
After a long time ($t=1000$), the flow distributions settle down
into a piecewise constant state which are shown in the fist picture of
fig.~\ref{iso-xvelo}, where the symbols are the numerical solutions
and the solid lines are the exact hydrostatic solutions. The
velocity distributions are also shown in fig.~\ref{iso-xvelo}. For the
$1^{st}$ order schemes, the oscillation of velocity around zero is
on the order of $10^{-7}$. This is mainly caused by the error in
numerical integrations because there is no exact solution for most
integrals in Eq.(\ref{flux1-l})-(\ref{flux2-r}). In fact, the
precision of numerical integration for the integrals is on the order
of $10^{-6}\sim10^{-7}$. Since the potential jumps are large and the
high order scheme uses more integral evaluations,  the velocity
distribution calculated by $2^{nd}$ order scheme is a little bit
worse than the 1st-order ones. If a better accuracy can be achieved
for the numerical evaluation of the integrals, the velocity error
can be further reduced to machine zero.

\subsection{Perturbation of the 1D isothermal equilibrium solution}

\begin{figure}
\begin{center}
\includegraphics[scale=0.35,bb = 80 40 710 570, clip=true]{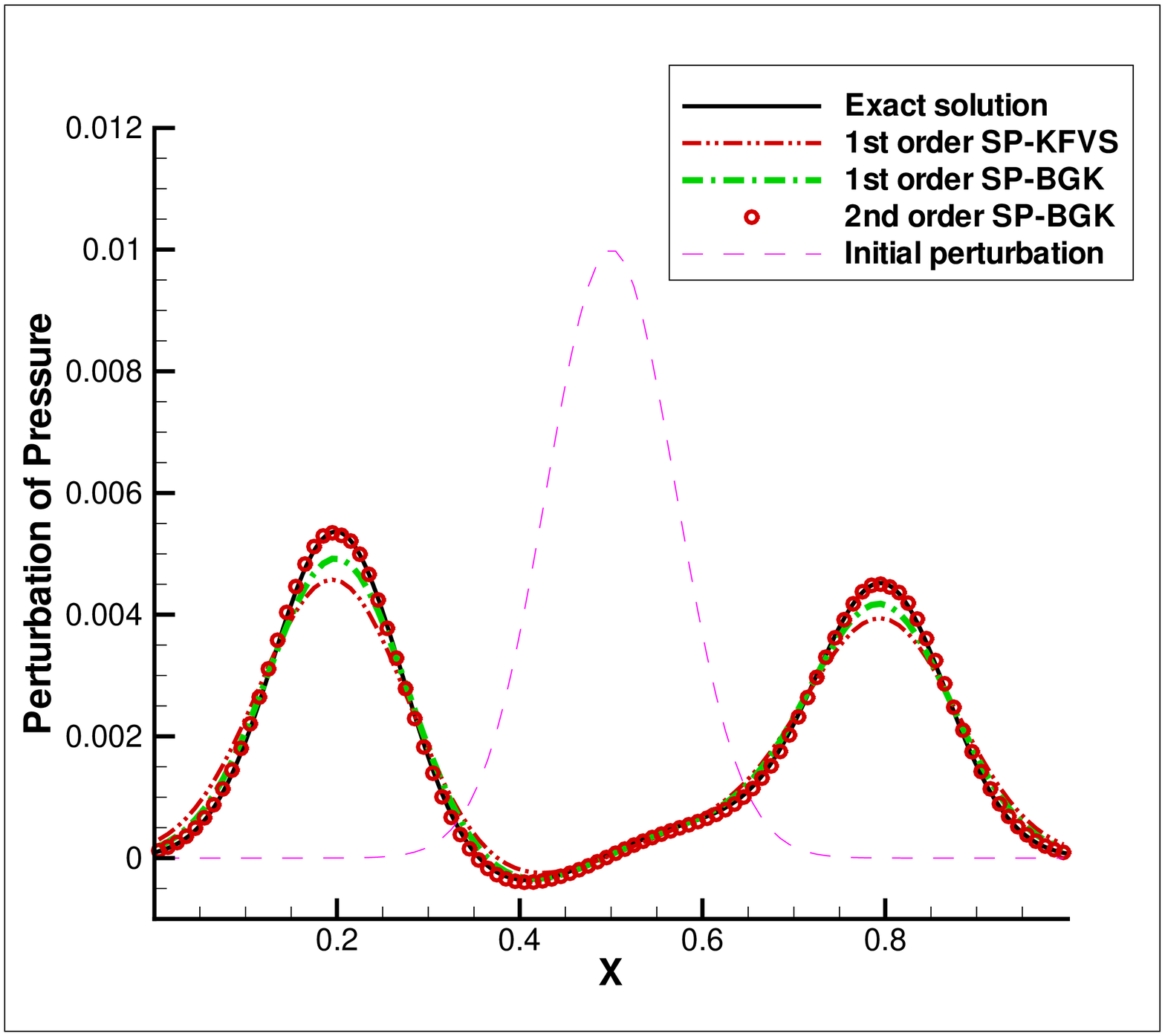}
\includegraphics[scale=0.35,bb = 80 40 715 575, clip=true]{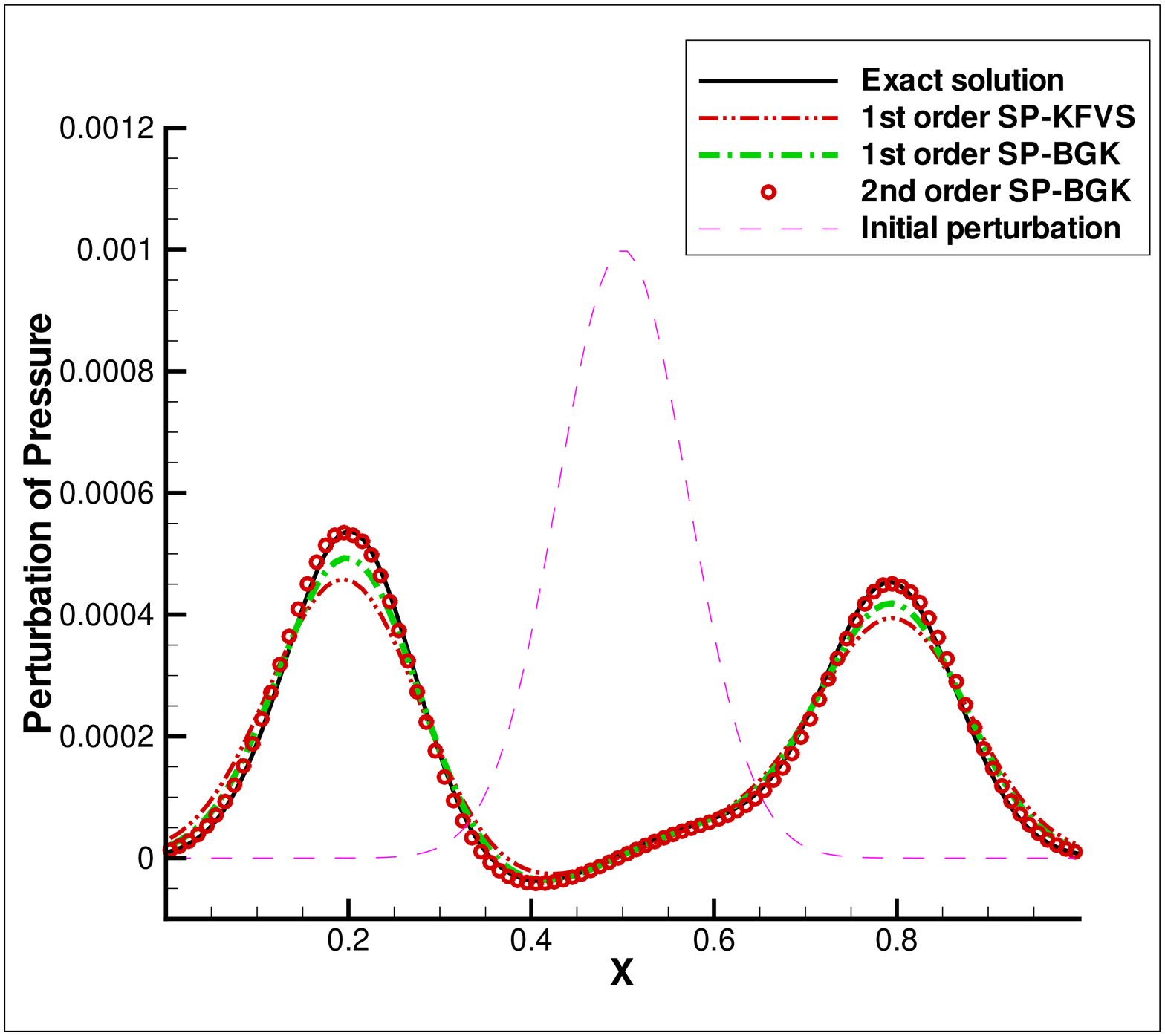}
\caption{\small Perturbation of pressure on an isothermal
equilibrium solution. Left: $\eta=0.01$; right: $\eta=0.001$.}\label{hydro}
\end{center}
\end{figure}

This test case is from LeVeque and Bale's paper \cite{leveque}. We
consider an ideal gas with $\gamma=1.4$ on an initial isothermal
hydrostatic state,
\[\rho_0(x)=p_0(x)=e^{-x},\textrm{ and }U_0(x)=0,\]
for $x \in [0,1]$. Initially, the pressure is perturbed by
\[p(x,t=0)=p_0(x)+\eta e^{\alpha (x-x_0)^2},\]
where $\alpha=100$, $x_0=0.5$ and $\eta$ is the amplitude of the
perturbation. The gravitational field is the same as in example
\ref{shock}. The computation is conducted with $100$ grid points in
the whole domain and stops at time $t=0.25$. Fig.~\ref{hydro},
show the results from SP-KFVS and SP-BGK schemes,
where SP-KFVS has larger numerical dissipation than SP-BGK scheme.
The results calculated by the $2^{nd}$ order SP-BGK scheme matches the
exact solution very well.

\begin{figure}
\begin{center}
\includegraphics[scale=0.35,bb = 80 35 700 560, clip=true]{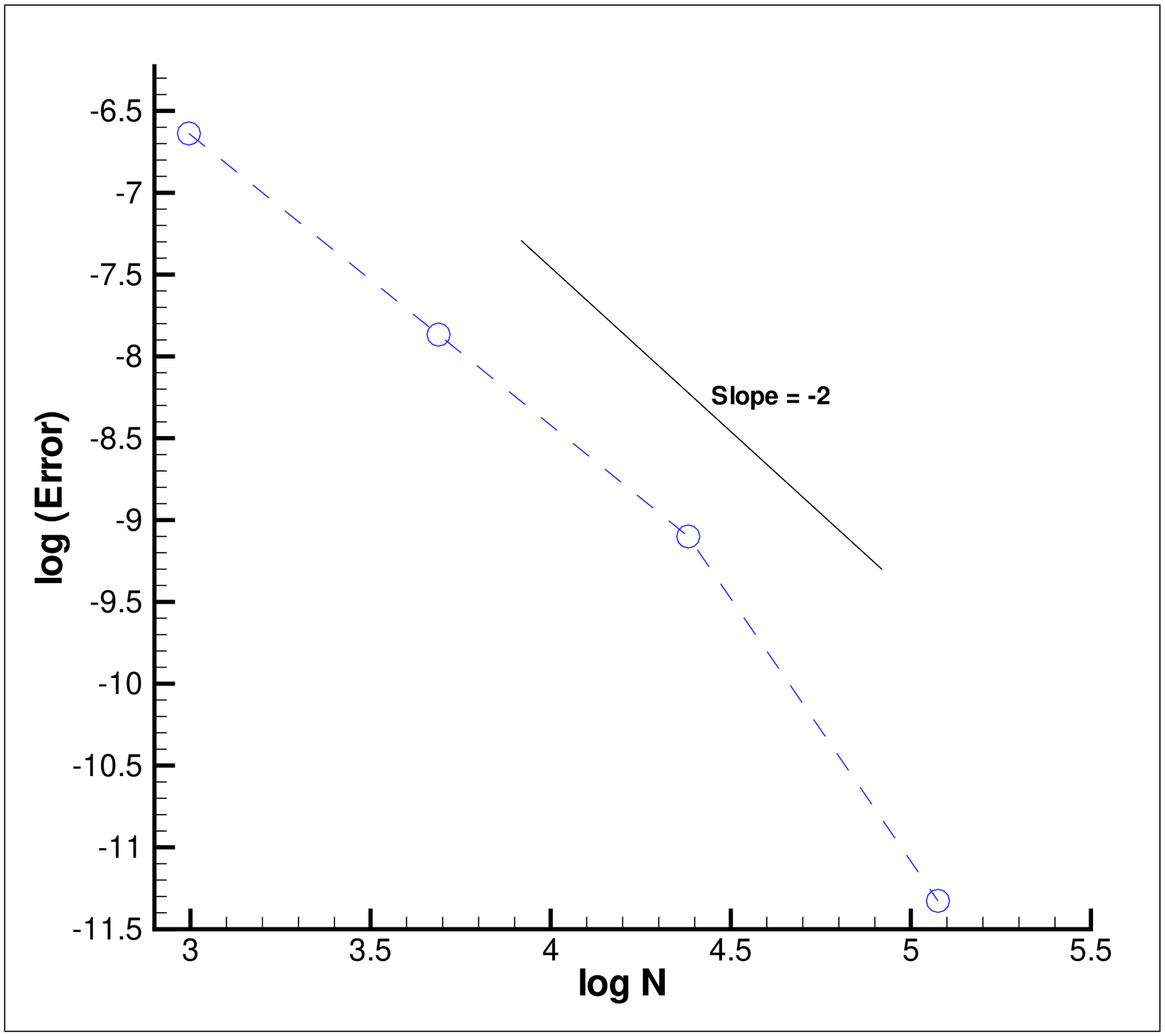}
\includegraphics[scale=0.35,bb = 80 35 700 560, clip=true]{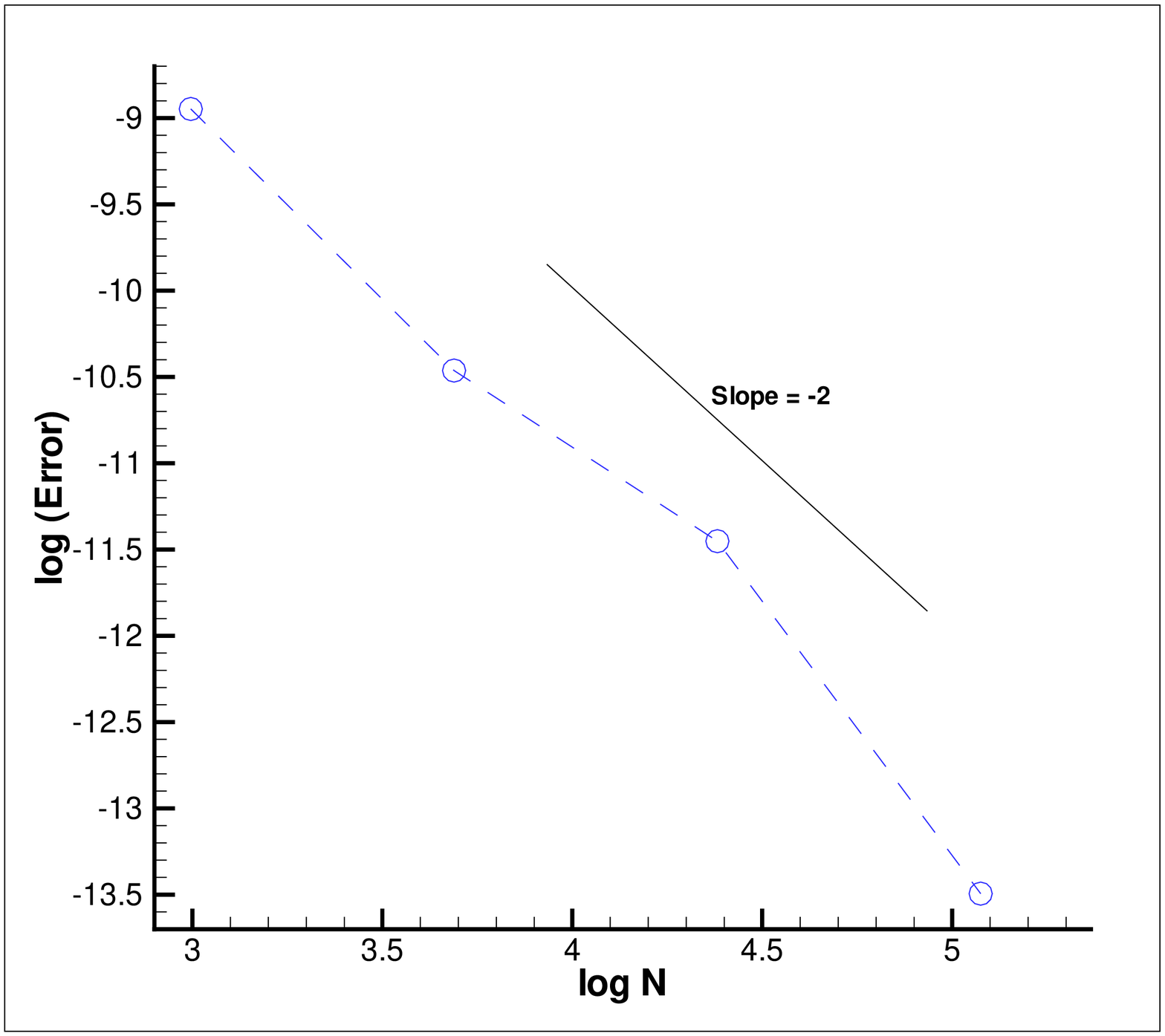}
\caption{\small Convergency rate of the $2^{nd}$-order SP-BGK scheme
for perturbation of pressure on an isothermal equilibrium solution
with $\eta=0.01$ on the left figure, and $\eta=0.001$ on the right
figure.}\label{error}
\end{center}
\end{figure}

Also in fig.~\ref{error}, we show the convergency rate of our
$2^{nd}$-order SP-BGK scheme, where the number of cells is N and the
error is the $L^\infty$ error. From the figures, we can conclude our
$2^{nd}$-order SP-BGK scheme has a 2nd-order accuracy  even with the
modeling of piecewise constant potential.

\subsection{One-dimension gas falling into a fixed external potential.}

\begin{figure}
\begin{center}
\includegraphics[scale=0.35,bb = 79 40 715 575, clip=true]{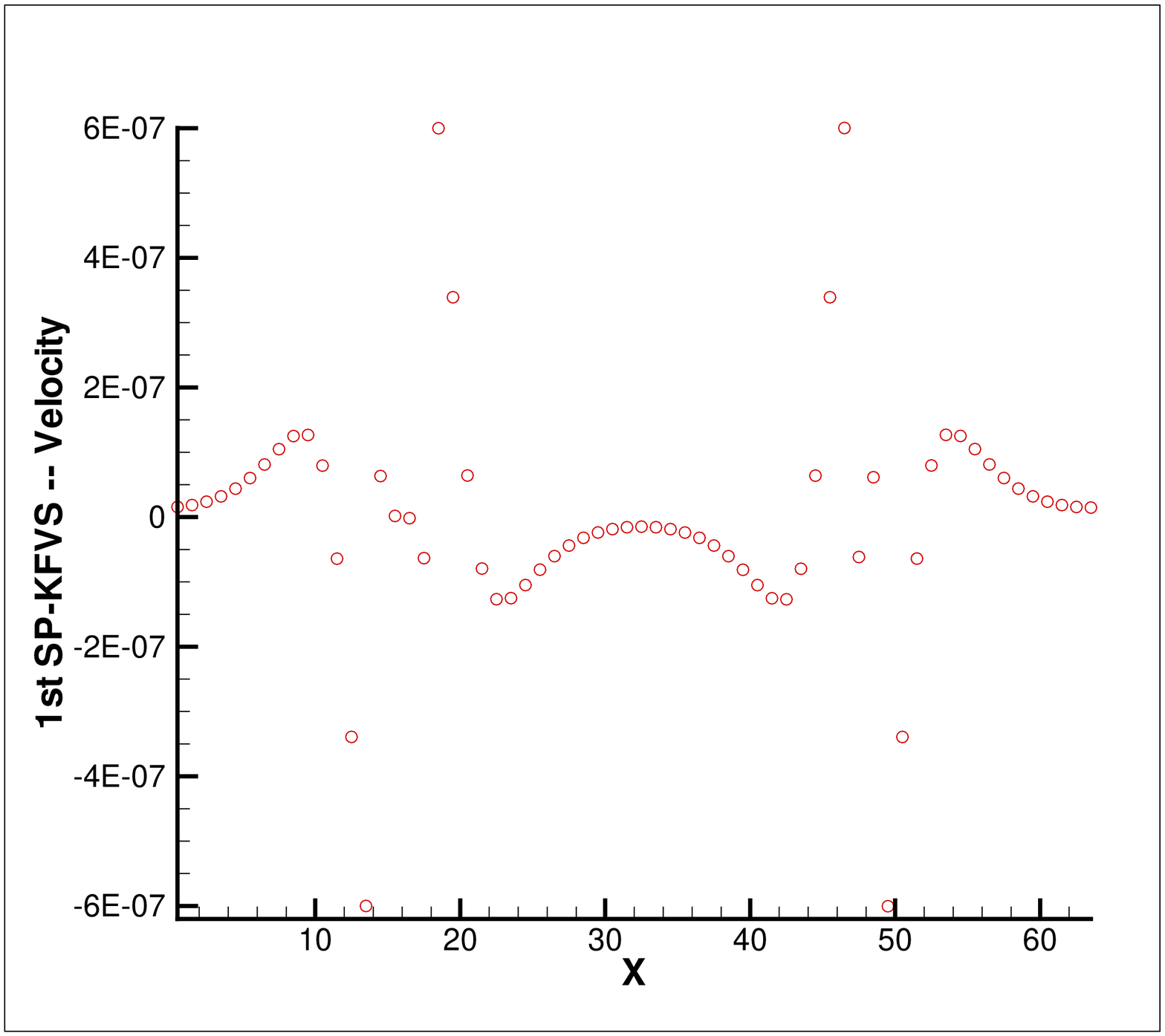}
\includegraphics[scale=0.35,bb = 79 40 715 585, clip=true]{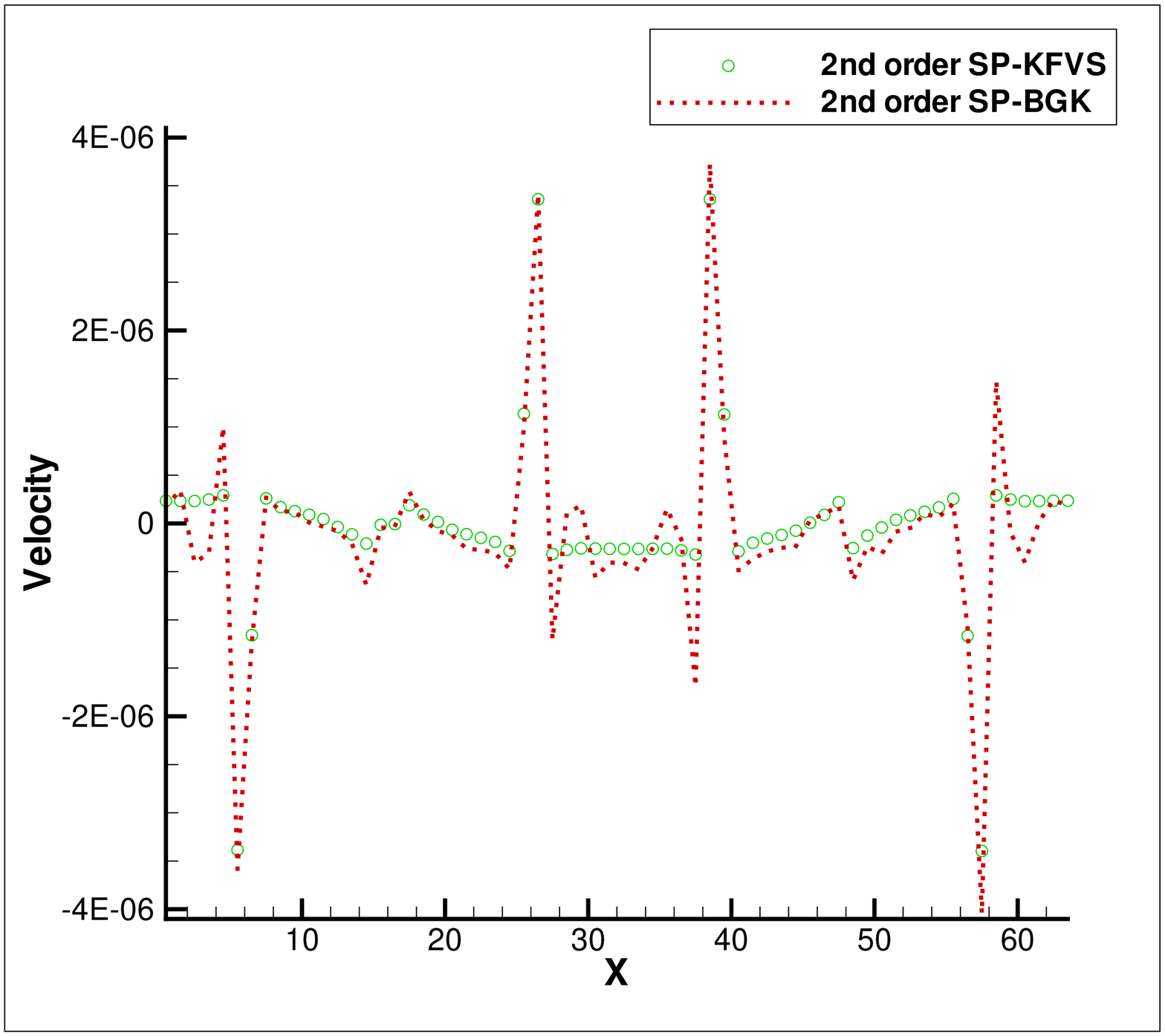}
\caption{\small Velocity distributions for gas falling into a fixed
external potential in 1-D case. The exact solution should have a
zero velocity. The error is due to the numerical integration, e.g.,
$\int_{-\infty}^0 g (u) (-\frac{u}{\sqrt{u^2+U_c^2}}) \ud u$, where
there is no analytic solution. }\label{pren_xvelo}
\end{center}
\end{figure}

\begin{figure}
\begin{center}
\includegraphics[scale=0.35,bb = 79 40 715 580, clip=true]{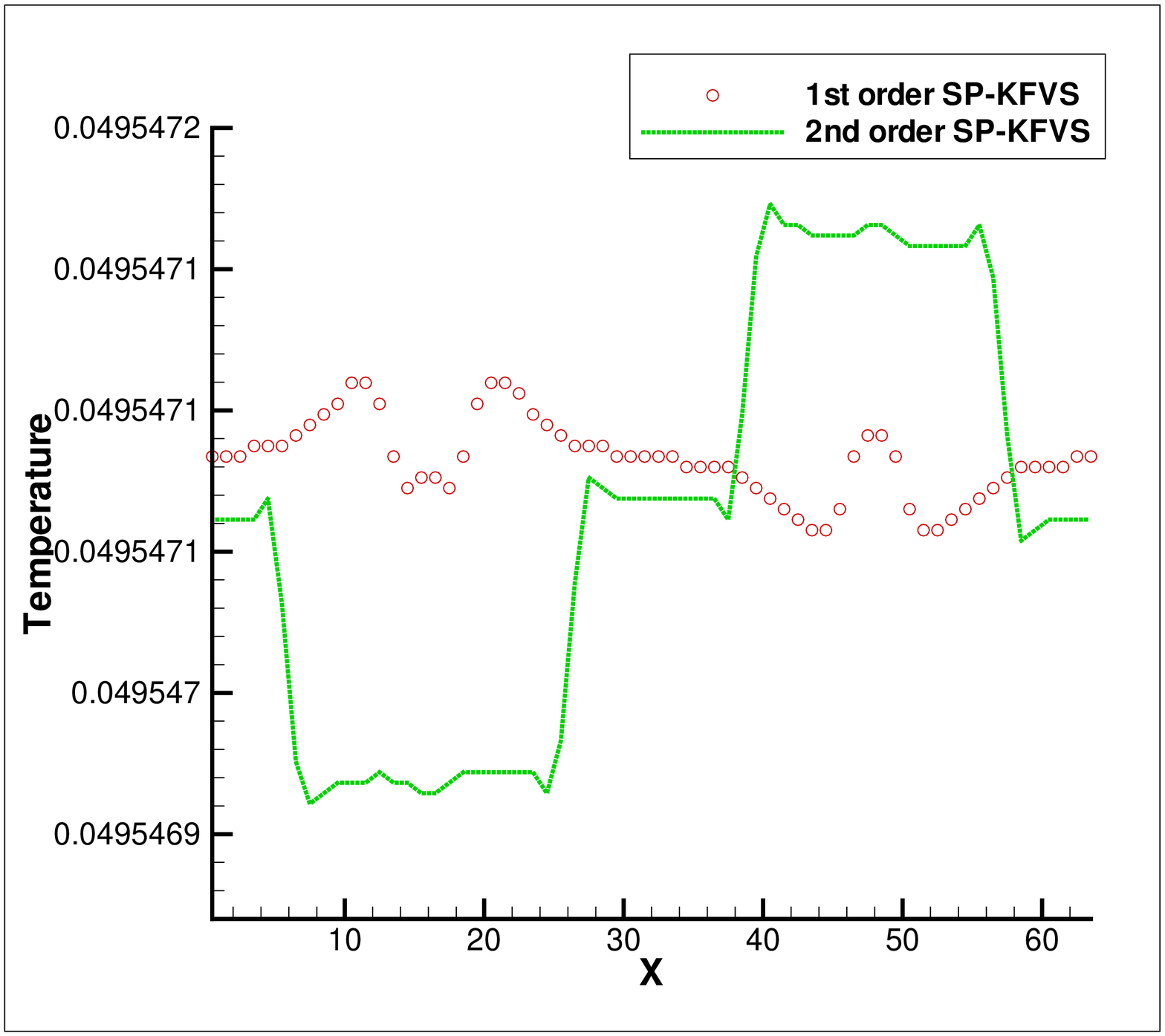}
\includegraphics[scale=0.35,bb = 79 40 715 580, clip=true]{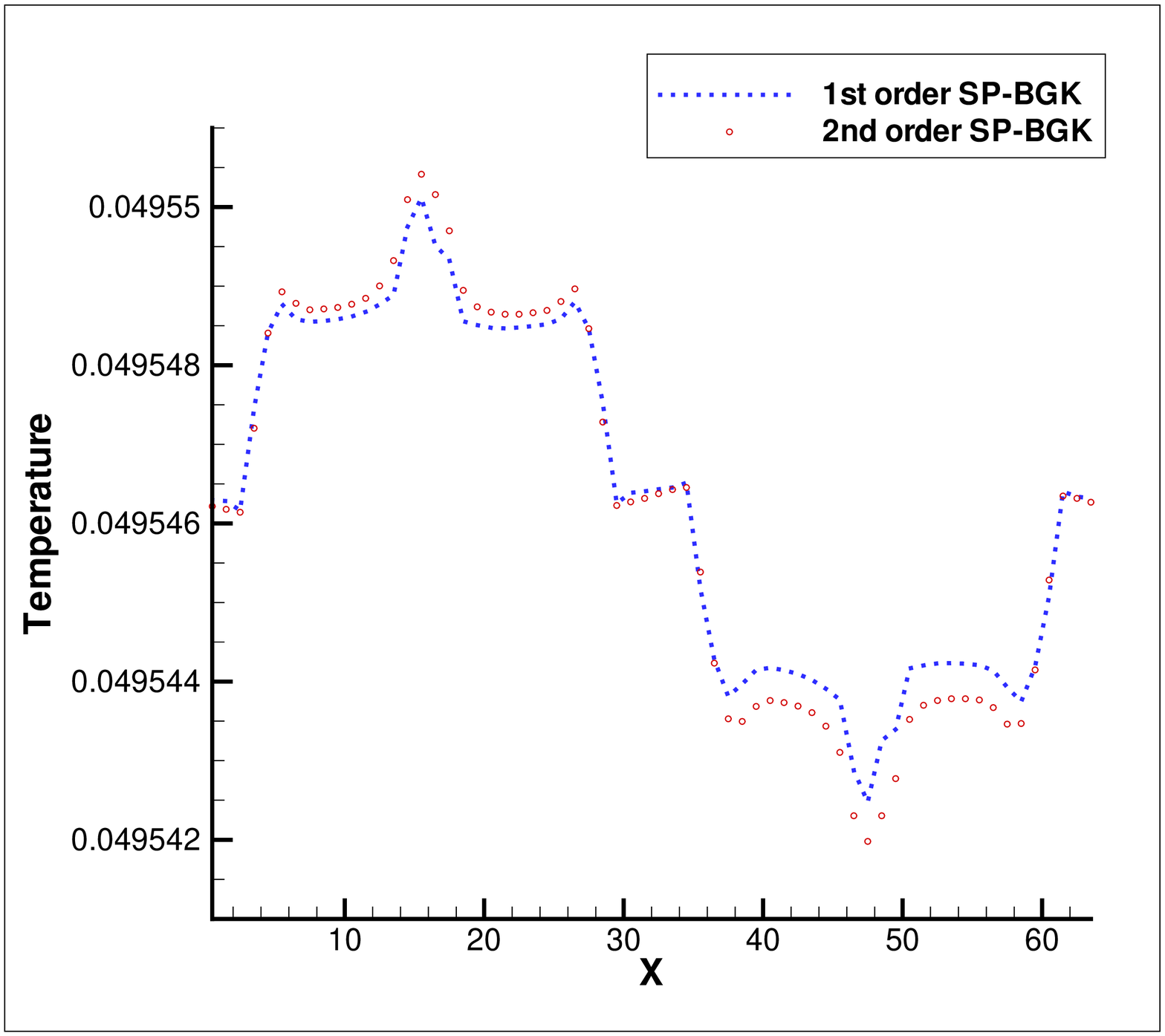}
\caption{\small Temperature distributions for gas falling into a
fixed external potential in 1-D case.}\label{pren_temp}
\end{center}
\end{figure}

This case is taken from the paper by Slyz and Prendergast \cite{slyz} to investigate the
numerical accuracy of the BGK scheme. The gas is initially stationary ($U=0$) and homogeneous
($\rho=1$, $e=1$, where $e$ is the internal energy). The gravitational potential has the form of a sine wave,
\[
\phi=-\phi_0\frac{L}{2\pi}\sin \frac{2\pi x}{L},
\]
where $L=64$ is the length of the computational domain and
$\phi_0=0.02$. The ratio of the specific heat $\gamma =5/3$. The
periodic boundary conditions are implemented in this system.
Simulation results are presented with $\Delta x=1$ and at the output
time $t=250000$ (more than 500000 time steps). After the initial
transition, the system is expected to reach an isothermal
hydrostatic distribution, where the temperature settles to a
constant with zero  velocity, i.e.,
\[T(x,t)=T_0,\textrm{ and }U=0.\]The velocity and temperature distributions computed by
different symplecticity preserving schemes are shown in
fig.~\ref{pren_xvelo}, ~\ref{pren_temp}. The numerical
error is smaller than that in \cite{tian}. Moreover, the results can be further improved if a better
numerical integration for the integral evaluation can be adopted.

\subsection{Rayleigh-Taylor instability.}
\begin{figure}
\begin{center}
\includegraphics[angle=270,scale=0.45,bb = 30 80 580 700, clip=true]{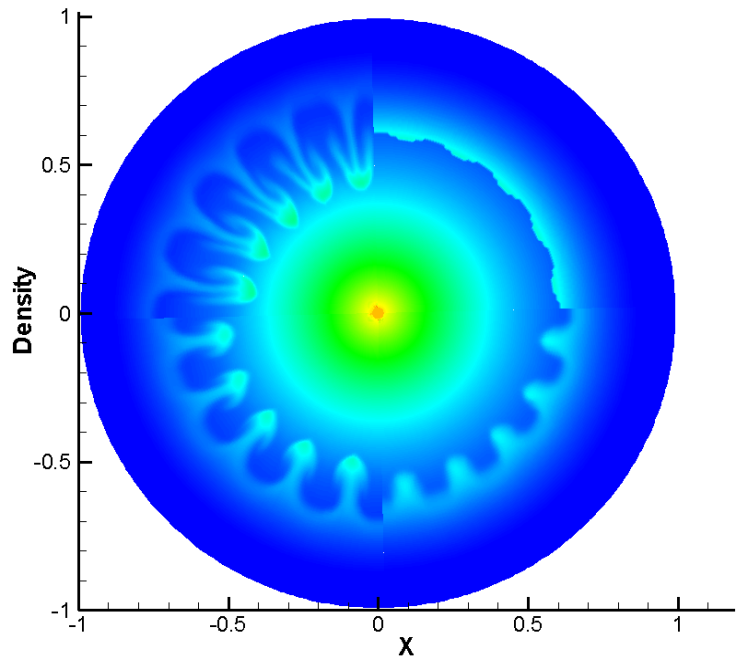}
\caption{\small Rayleigh-Taylor instability with gravitational field
directed radially inward. Density contours at time $t=0, 0.8, 1.4,
2.0$ are shown in the four quadrants, starting with the initial data
in the upper right corner and progressing clockwise.}\label{r-t}
\end{center}
\end{figure}

\begin{figure}
\begin{center}
\includegraphics[scale=0.35,bb = 80 40 700 570, clip=true]{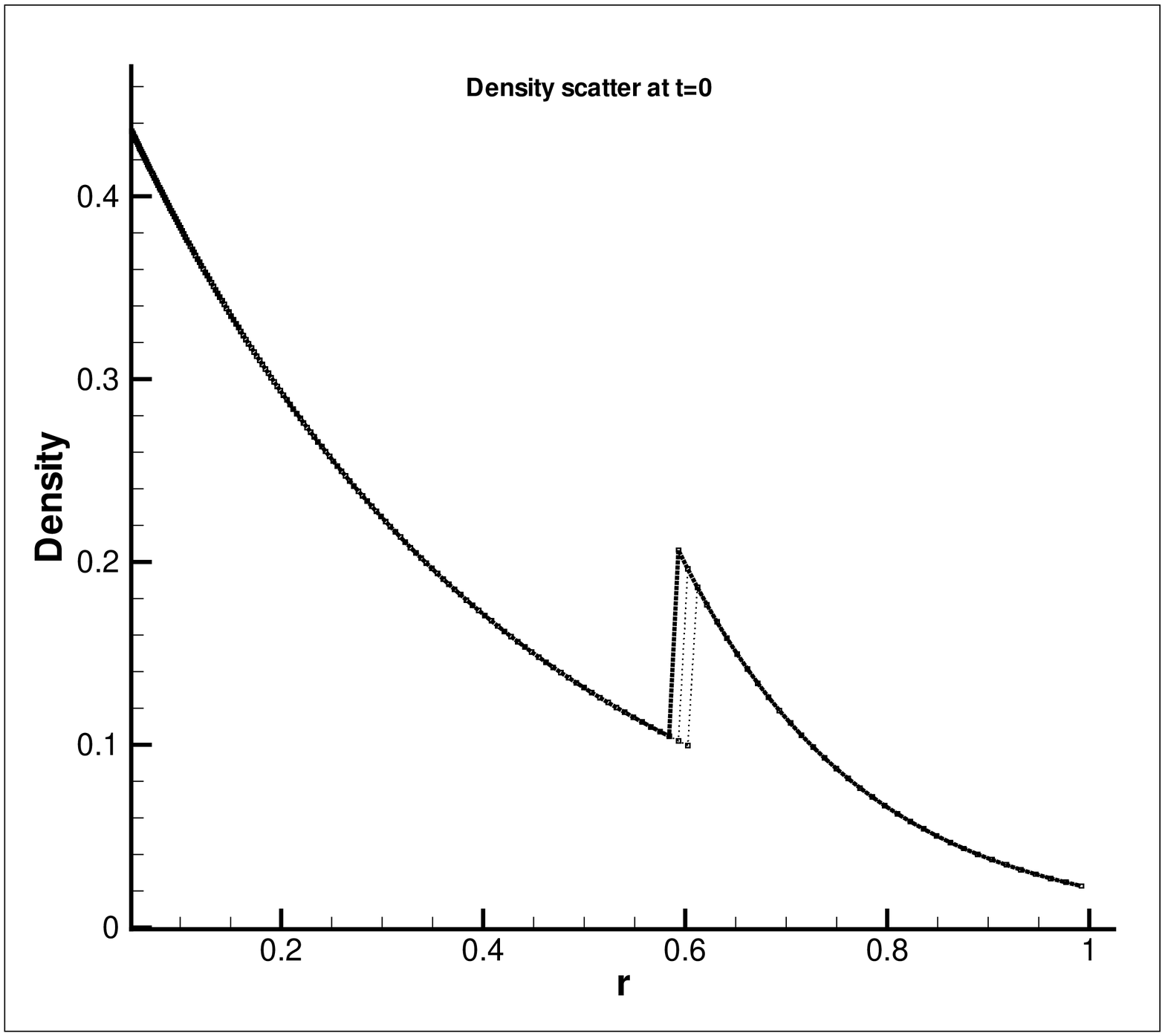}
\includegraphics[scale=0.35,bb = 90 40 700 570, clip=true]{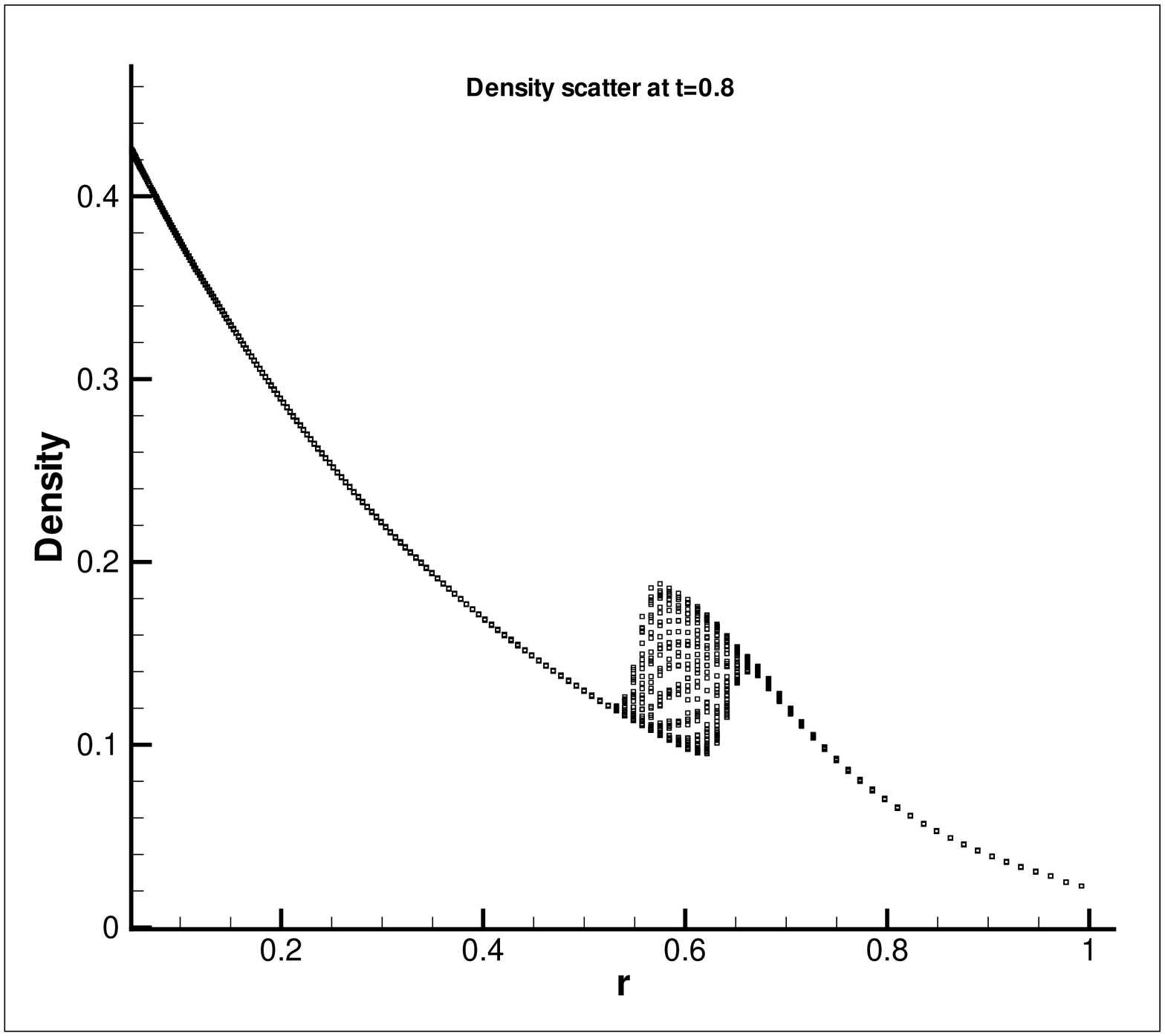}
\includegraphics[scale=0.35,bb = 80 40 700 570, clip=true]{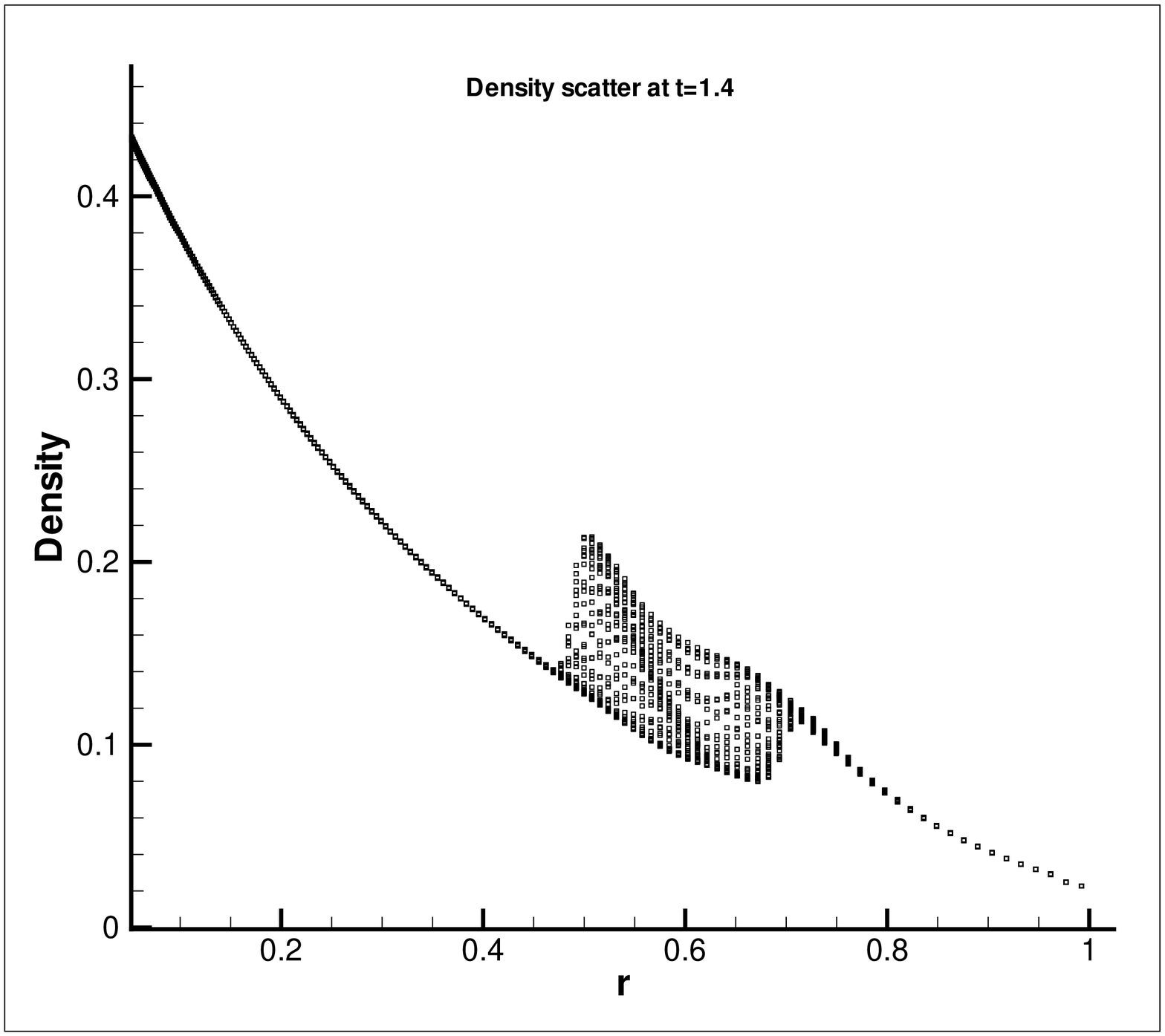}
\includegraphics[scale=0.35,bb = 80 40 700 570, clip=true]{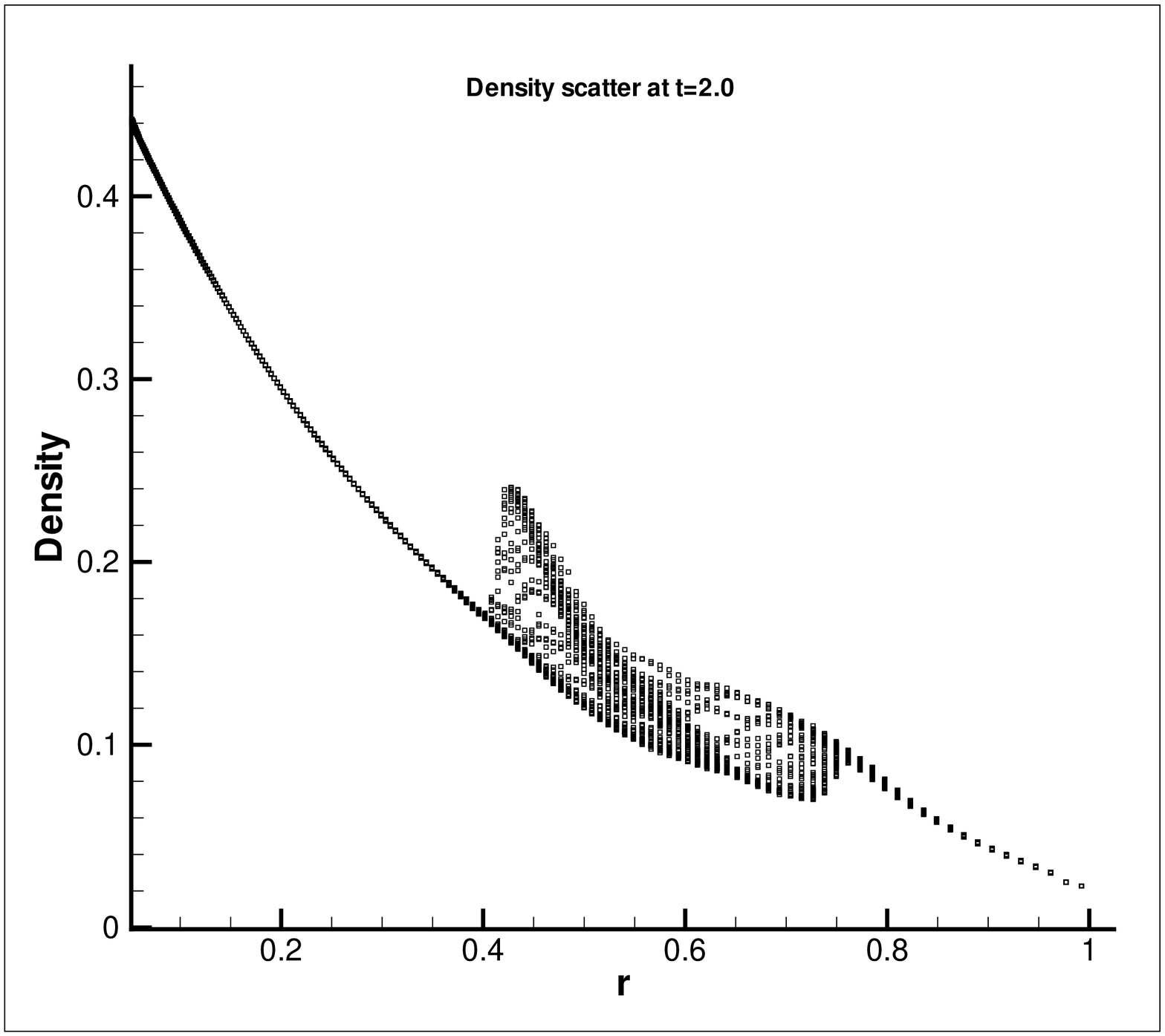}
\caption{\small Scatter plots of the density in the cell vs. the distance of the cell center from the origin.}\label{taylor-1d}
\end{center}
\end{figure}

This test case also comes from \cite{leveque}. Consider an isothermal equilibrium idea gas ($\gamma=1.4$) in a 2D polar coordinate $(r,\theta)$,
\[
 \rho_0(r)=e^{-\alpha(r+r_0)},\, p_0(r)=\frac{1.5}{\alpha}e^{-\alpha(r+r_0)},\,U_0=0,
\]
where
\[
\left\{\begin{array}{lr}
\alpha=2.68,\,r_0=0.258\quad & \text{for}\, r\le r_1,\\ \\
\alpha=5.53,\,r_0=-0.308\quad & \text{for}\, r> r_1,
\end{array}
\right.
\,\text{and}\,
\left\{\begin{array}{lr}
r_1=0.6(1+0.02\cos(20\theta))\quad & \text{for}\, density,\\ \\
r_1=0.62324965 \quad & \text{for}\, pressure,
\end{array}
\right.
\]
The potential satisfies $-\nabla\phi(r)=1.5$.  The time evolutions
of the density distributions at times $t=0, 0.8, 1.4$ and $2.0$ are
shown in fig.~\ref{r-t}. Fig.~\ref{taylor-1d} shows a scatter plot
of the density as a function of the radius. These figures clearly
show that the hydrostatic solution can be well kept and the flow
motion is limited around the unstable interface.

\section{Conclusion}\label{sec6}

In this paper, based on the Liouville's theorem and
symplecticity-preserving property of a Hamiltonian flow, a
well-balanced gas-kinetic BGK scheme (SP-BGK) has been developed for
a hydrodynamic system under gravitational field with the modeling of
piecewise constant potentials. As shown in the paper, in order to
design such a scheme, the equilibrium state used has to be an exact
Maxwellian distribution function. At the same time, the physical
mechanism of particle transport across a potential barrier has to be
explicitly followed in the equilibrium states modeling and the flux
evaluation. As far as we know, the method presented in this paper is
the first exact well-balanced scheme for the Navier-Stokes equations
under gravitational field. At the same time, the particle transport
mechanism across a potential jump in the current kinetic formulation
follows the physical principles closely, which is valid under any
general physical situation. Both the shock capturing and
well-balanced properties are automatically obtained under the
corresponding physical conditions. Mathematically, it has been
proved that the SP-BGK method is a well-balanced scheme which could
keep the hydrostatic state forever. In this paper, the design of the
well-balanced scheme comes from the first principles of physics,
instead of using the well-balanced condition as the starting point
in the design of such a scheme.\\ \\

\section*{Acknowledgments}

The current research  was supported by Hong Kong Research Grant
Council 621709, National Natural Science Foundation of China
(Project No. 10928205), National Key Basic Research Program
(2009CB724101).

\bigskip

\section*{Appendix}

{\bf Formulae in the two-dimensional case:}\\

1. Equilibrium states\\

Case 1. $\phi_j<\phi_{j+1}$, define
$U_c=\sqrt{2(\phi_{j+1}-\phi_j)}$.
\begin{equation}\label{var1-2l}
\begin{array}{l}
\quad W_{j+1/2}^l
= \iiint^{+\infty}_0 f_j(x_{j+1/2},0,u,v,\xi) \left(
\begin{array}{c}
1\\
u\\
v\\
\frac{1}{2}(u^2+v^2+\xi^2)
\end{array} \right)
\ud u \,\ud v \,\ud \xi\\ \\
\quad +\iiint^{U_c}_0 f_j(x_{j+1/2},0,u,v,\xi) \left(
\begin{array}{c}
1\\
-u\\
v\\
\frac{1}{2}(u^2+\xi^2)
\end{array} \right)
\ud u \,\ud v \,\ud \xi\\ \\
\quad +\iiint_{-\infty}^0 f_{j+1}(x_{j+1/2},0,u,v,\xi) \left(
\begin{array}{c}
-\frac{u}{\sqrt{u^2+v^2+U_c^2}}\\
u\\
-\frac{uv}{\sqrt{u^2+U_c^2}}\\
\frac{1}{2}(-u\sqrt{u^2+U_c^2}-\frac{uv^2}{\sqrt{u^2+U_c^2}}-\frac{u}{\sqrt{u^2+U_c^2}}\xi^2)
\end{array} \right)
\ud u \,\ud v \, \ud \xi.
\end{array}
\end{equation}

\begin{equation}\label{var1-2r}
\begin{array}{l}
W_{j+1/2}^r= \iiint^{+\infty}_{U_c} f_j(x_{j+1/2},0,u,v,\xi) \left(
\begin{array}{c}
\frac{u}{\sqrt{u^2-U_c^2}}\\
u\\
\frac{uv}{\sqrt{u^2-U_c^2}}\\
\frac{1}{2}(u\sqrt{u^2-U_c^2}+\frac{uv^2}{\sqrt{u^2-U_c^2}}+\frac{u}{\sqrt{u^2-U_c^2}}\xi^2)
\end{array} \right)
\ud u\,\ud v \,\ud \xi\\ \\
\quad +\iiint_{-\infty}^0 f_{j+1}(x_{j+1/2},0,u,v,\xi) \left(
\begin{array}{c}
1\\
u\\
v\\
\frac{1}{2}(u^2+v^2+\xi^2)
\end{array} \right)
\ud u \, \ud v\,\ud \xi.
\end{array}
\end{equation}\\

Case 2. $\phi_j>\phi_{j+1}$, define
$U_c=\sqrt{2(\phi_j-\phi_{j+1})}$.
\begin{equation}\label{var2-2l}
\begin{array}{l}
W_{j+1/2}^l= \iiint^{+\infty}_0 f_j(x_{j+1/2},0,u,\xi) \left(
\begin{array}{c}
1\\
u\\
v\\
\frac{1}{2}(u^2+v^2+\xi^2)
\end{array} \right)
\ud u \,\ud v\, \ud \xi\\ \\
\quad+\iiint_{-\infty}^{-U_c} f_{j+1}(x_{j+1/2},0,u,\xi)
\left( \begin{array}{c}
-\frac{u}{\sqrt{u^2-U_c^2}}\\
u\\
-\frac{uv}{\sqrt{u^2-U_c^2}}\\
\frac{1}{2}(-u\sqrt{u^2-U_c^2}-\frac{uv^2}{\sqrt{u^2-U_c^2}}-\frac{u}{\sqrt{u^2-U_c^2}}\xi^2)
\end{array} \right)
\ud u \,\ud v\,\ud \xi.
\end{array}
\end{equation}

\begin{equation}\label{var2-2r}
\begin{array}{l}
W_{j+1/2}^r= \iiint^{+\infty}_0 f_j(x_{j+1/2},0,u,\xi) \left(
\begin{array}{c}
\frac{u}{\sqrt{u^2+U_c^2}}\\
u\\
\frac{uv}{\sqrt{u^2+U_c^2}}\\
\frac{1}{2}(u\sqrt{u^2+U_c^2}+\frac{uv^2}{\sqrt{u^2+U_c^2}}+\frac{u}{\sqrt{u^2+U_c^2}}\xi^2)
\end{array} \right)
\ud u \,\ud v\, \ud \xi\\ \\
\quad +\iiint_{-U_c}^0 f_{j+1}(x_{j+1/2},0,u,\xi) \left(
\begin{array}{c}
1\\
-u\\
v\\
\frac{1}{2}(u^2+v^2+\xi^2)
\end{array} \right)
\ud u \,\ud v\, \ud \xi\\ \\
\quad +\iiint_{-\infty}^0 f_{j+1}(x_{j+1/2},0,u,\xi) \left(
\begin{array}{c}
1\\
u\\
v\\
\frac{1}{2}(u^2+v^2+\xi^2)
\end{array} \right)
\ud u \,\ud v\, \ud \xi.
\end{array}
\end{equation}

2. Fluxes\\

Case 1. $\phi_j<\phi_{j+1}$, define $U_c=\sqrt{2(\phi_{j+1}-\phi_j)}$.
\begin{equation}\label{flux1-2l}
\begin{array}{l}
F^l_{j+1/2}(t)= \iiint^{+\infty}_0 f_j(x_{j+1/2},t,u,\xi)
\left( \begin{array}{c}
u\\
u^2\\
uv\\
\frac{1}{2}(u^3+uv^2+u\xi^2)
\end{array} \right)
\ud u \,\ud v\, \ud \xi\\ \\
\qquad \qquad \quad+\iiint^{U_c}_0 f_j(x_{j+1/2},t,u,\xi)
\left( \begin{array}{c}
-u\\
u^2\\
-uv\\
\frac{1}{2}(-u^3-uv^2-u\xi^2)
\end{array} \right)
\ud u \,\ud v\, \ud \xi\\ \\
\qquad \qquad \quad+\iiint_{-\infty}^0 f_{j+1}(x_{j+1/2},t,u,\xi)
\left( \begin{array}{c}
u\\
-u\sqrt{u^2+U_c^2}\\
uv\\
\frac{1}{2}(u(u^2+U_c^2)+uv^2+u\xi^2)
\end{array} \right)
\ud u \,\ud v\, \ud \xi.
\end{array}
\end{equation}

\begin{equation}\label{flux1-2r}
\begin{array}{l}
F^r_{j+1/2}(t)= \iiint^{+\infty}_{U_c} f_j(x_{j+1/2},t,u,\xi)
\left( \begin{array}{c}
u\\
u\sqrt{u^2-U_c^2}\\
uv\\
\frac{1}{2}(u(u^2-U_c^2)+uv^2+u\xi^2)
\end{array} \right)
\ud u \,\ud v\, \ud \xi\\ \\
\qquad \qquad \quad+\iiint_{-\infty}^0 f_{j+1}(x_{j+1/2},t,u,\xi)
\left( \begin{array}{c}
u\\
u^2\\
uv\\
\frac{1}{2}(u^3+uv^2+u\xi^2)
\end{array} \right)
\ud u \,\ud v\, \ud \xi.
\end{array}
\end{equation}\\

Case 2. $\phi_j>\phi_{j+1}$, define $U_c=\sqrt{2(\phi_j-\phi_{j+1})}$.
\begin{equation}\label{flux2-2l}
\begin{array}{l}
F^l_{j+1/2}(t)= \iiint^{+\infty}_0 f_j(x_{j+1/2},t,u,\xi)
\left( \begin{array}{c}
u\\
u^2\\
uv\\
\frac{1}{2}(u^3+uv^2+u\xi^2)
\end{array} \right)
\ud u \,\ud v\, \ud \xi\\ \\
\qquad \qquad \quad+\iiint_{-\infty}^{-U_c} f_{j+1}(x_{j+1/2},t,u,\xi)
\left( \begin{array}{c}
u\\
-u\sqrt{u^2-U_c^2}\\
uv\\
\frac{1}{2}(u(u^2-U_c^2)+uv^2+u\xi^2)
\end{array} \right)
\ud u \,\ud v\, \ud \xi.
\end{array}
\end{equation}

\begin{equation}\label{flux2-2r}
\begin{array}{l}
F^r_{j+1/2}(t)= \iiint^{+\infty}_0 f_j(x_{j+1/2},t,u,\xi)
\left( \begin{array}{c}
u\\
u\sqrt{u^2+U_c^2}\\
uv\\
\frac{1}{2}(u(u^2+U_c^2)+uv^2+u\xi^2)
\end{array} \right)
\ud u \,\ud v\, \ud \xi\\ \\
\qquad \qquad \quad+\iiint_{-U_c}^0 f_{j+1}(x_{j+1/2},t,u,\xi)
\left( \begin{array}{c}
-u\\
u^2\\
-uv\\
\frac{1}{2}(-u^3-uv^2-u\xi^2)
\end{array} \right)
\ud u \,\ud v\, \ud \xi\\ \\
\qquad \qquad \quad+\iiint_{-\infty}^0 f_{j+1}(x_{j+1/2},t,u,\xi)
\left( \begin{array}{c}
u\\
u^2\\
uv\\
\frac{1}{2}(u^3+uv^2+u\xi^2)
\end{array} \right)
\ud u \,\ud v\, \ud \xi.
\end{array}
\end{equation}\\

{\bf Formulae in the one-dimensional case:}\\

1. Equilibrium states: \\

Case 1. $\phi_j<\phi_{j+1}$, define
$U_c=\sqrt{2(\phi_{j+1}-\phi_j)}$.
\begin{equation}\label{var1-l}
\begin{array}{l}
W_{j+1/2}^l= \iint^{+\infty}_0 f_j(x_{j+1/2},0,u,\xi) \left(
\begin{array}{c}
1\\
u\\
\frac{1}{2}(u^2+\xi^2)
\end{array} \right)
\ud u \, \ud \xi\\ \\
\qquad \qquad +\iint^{U_c}_0 f_j(x_{j+1/2},0,u,\xi) \left(
\begin{array}{c}
1\\
-u\\
\frac{1}{2}(u^2+\xi^2)
\end{array} \right)
\ud u \, \ud \xi\\ \\
\qquad \qquad +\iint_{-\infty}^0 f_{j+1}(x_{j+1/2},0,u,\xi) \left(
\begin{array}{c}
-\frac{u}{\sqrt{u^2+U_c^2}}\\
u\\
\frac{1}{2}(-u\sqrt{u^2+U_c^2}-\frac{u}{\sqrt{u^2+U_c^2}}\xi^2)
\end{array} \right)
\ud u \, \ud \xi.
\end{array}
\end{equation}

\begin{equation}\label{var1-r}
\begin{array}{l}
W_{j+1/2}^r= \iint^{+\infty}_{U_c} f_j(x_{j+1/2},0,u,\xi) \left(
\begin{array}{c}
\frac{u}{\sqrt{u^2-U_c^2}}\\
u\\
\frac{1}{2}(u\sqrt{u^2-U_c^2}+\frac{u}{\sqrt{u^2-U_c^2}}\xi^2)
\end{array} \right)
\ud u \, \ud \xi\\ \\
\qquad \qquad +\iint_{-\infty}^0 f_{j+1}(x_{j+1/2},0,u,\xi) \left(
\begin{array}{c}
1\\
u\\
\frac{1}{2}(u^2+\xi^2)
\end{array} \right)
\ud u \, \ud \xi.
\end{array}
\end{equation}\\

Case 2. $\phi_j>\phi_{j+1}$, define
$U_c=\sqrt{2(\phi_j-\phi_{j+1})}$.
\begin{equation}\label{var2-l}
\begin{array}{l}
W_{j+1/2}^l= \iint^{+\infty}_0 f_j(x_{j+1/2},0,u,\xi) \left(
\begin{array}{c}
1\\
u\\
\frac{1}{2}(u^2+\xi^2)
\end{array} \right)
\ud u \, \ud \xi\\ \\
\qquad \qquad+\iint_{-\infty}^{-U_c} f_{j+1}(x_{j+1/2},0,u,\xi)
\left( \begin{array}{c}
-\frac{u}{\sqrt{u^2-U_c^2}}\\
u\\
\frac{1}{2}(-u\sqrt{u^2-U_c^2}-\frac{u}{\sqrt{u^2-U_c^2}}\xi^2)
\end{array} \right)
\ud u \, \ud \xi.
\end{array}
\end{equation}

\begin{equation}\label{var2-r}
\begin{array}{l}
W_{j+1/2}^r= \iint^{+\infty}_0 f_j(x_{j+1/2},0,u,\xi) \left(
\begin{array}{c}
\frac{u}{\sqrt{u^2+U_c^2}}\\
u\\
\frac{1}{2}(u\sqrt{u^2+U_c^2}+\frac{u}{\sqrt{u^2+U_c^2}}\xi^2)
\end{array} \right)
\ud u \, \ud \xi\\ \\
\qquad \qquad +\iint_{-U_c}^0 f_{j+1}(x_{j+1/2},0,u,\xi) \left(
\begin{array}{c}
1\\
-u\\
\frac{1}{2}(u^2+\xi^2)
\end{array} \right)
\ud u \, \ud \xi\\ \\
\qquad \qquad +\iint_{-\infty}^0 f_{j+1}(x_{j+1/2},0,u,\xi) \left(
\begin{array}{c}
1\\
u\\
\frac{1}{2}(u^2+\xi^2)
\end{array} \right)
\ud u \, \ud \xi.
\end{array}
\end{equation}

2. Fluxes: \\

Case 1. $\phi_j<\phi_{j+1}$, define $U_c=\sqrt{2(\phi_{j+1}-\phi_j)}$.
\begin{equation}\label{flux1-l}
\begin{array}{l}
F^l_{j+1/2}(t)= \iint^{+\infty}_0 f_j(x_{j+1/2},t,u,\xi)
\left( \begin{array}{c}
u\\
u^2\\
\frac{1}{2}(u^3+u\xi^2)
\end{array} \right)
\ud u \, \ud \xi\\ \\
\qquad \qquad \quad+\iint^{U_c}_0 f_j(x_{j+1/2},t,u,\xi)
\left( \begin{array}{c}
-u\\
u^2\\
\frac{1}{2}(-u^3-u\xi^2)
\end{array} \right)
\ud u \, \ud \xi\\ \\
\qquad \qquad \quad+\iint_{-\infty}^0 f_{j+1}(x_{j+1/2},t,u,\xi)
\left( \begin{array}{c}
u\\
-u\sqrt{u^2+U_c^2}\\
\frac{1}{2}(u(u^2+U_c^2)+u\xi^2)
\end{array} \right)
\ud u \, \ud \xi.
\end{array}
\end{equation}

\begin{equation}\label{flux1-r}
\begin{array}{l}
F^r_{j+1/2}(t)= \iint^{+\infty}_{U_c} f_j(x_{j+1/2},t,u,\xi)
\left( \begin{array}{c}
u\\
u\sqrt{u^2-U_c^2}\\
\frac{1}{2}(u(u^2-U_c^2)+u\xi^2)
\end{array} \right)
\ud u \, \ud \xi\\ \\
\qquad \qquad \quad+\iint_{-\infty}^0 f_{j+1}(x_{j+1/2},t,u,\xi)
\left( \begin{array}{c}
u\\
u^2\\
\frac{1}{2}(u^3+u\xi^2)
\end{array} \right)
\ud u \, \ud \xi.
\end{array}
\end{equation}\\

Case 2. $\phi_j>\phi_{j+1}$, define $U_c=\sqrt{2(\phi_j-\phi_{j+1})}$.
\begin{equation}\label{flux2-l}
\begin{array}{l}
F^l_{j+1/2}(t)= \iint^{+\infty}_0 f_j(x_{j+1/2},t,u,\xi)
\left( \begin{array}{c}
u\\
u^2\\
\frac{1}{2}(u^3+u\xi^2)
\end{array} \right)
\ud u \, \ud \xi\\ \\
\qquad \qquad \quad+\iint_{-\infty}^{-U_c} f_{j+1}(x_{j+1/2},t,u,\xi)
\left( \begin{array}{c}
u\\
-u\sqrt{u^2-U_c^2}\\
\frac{1}{2}(u(u^2-U_c^2)+u\xi^2)
\end{array} \right)
\ud u \, \ud \xi.
\end{array}
\end{equation}

\begin{equation}\label{flux2-r}
\begin{array}{l}
F^r_{j+1/2}(t)= \iint^{+\infty}_0 f_j(x_{j+1/2},t,u,\xi)
\left( \begin{array}{c}
u\\
u\sqrt{u^2+U_c^2}\\
\frac{1}{2}(u(u^2+U_c^2)+u\xi^2)
\end{array} \right)
\ud u \, \ud \xi\\ \\
\qquad \qquad \quad+\iint_{-U_c}^0 f_{j+1}(x_{j+1/2},t,u,\xi)
\left( \begin{array}{c}
-u\\
u^2\\
\frac{1}{2}(-u^3-u\xi^2)
\end{array} \right)
\ud u \, \ud \xi\\ \\
\qquad \qquad \quad+\iint_{-\infty}^0 f_{j+1}(x_{j+1/2},t,u,\xi)
\left( \begin{array}{c}
u\\
u^2\\
\frac{1}{2}(u^3+u\xi^2)
\end{array} \right)
\ud u \, \ud \xi.
\end{array}
\end{equation}\\

\noindent{\bf Remarks on the integral evaluation:} in the above
formulae, there are many integrals which can not be analytically
evaluated, e.g., $\int_{-\infty}^0
f_{j+1}(-\frac{u}{\sqrt{u^2+U_c^2}}) \ud u$. Therefore, a numerical
integration method in \cite{recipes} has been used.

\newpage


\begin{figure}
\begin{center}
\includegraphics[scale=0.35,bb = 100 40 700 550, clip=true]{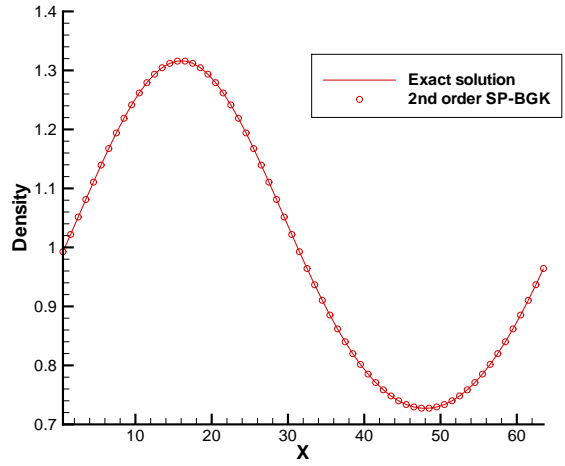}
\caption{\small Density distribution calculated by $2^{nd}$-order
SP-BGK for gas falling into a fixed external potential in 1-D
case.}\label{pren-dens}
\end{center}
\end{figure}

\end{document}